\documentclass[aps,prd,amsfonts,amssymb, amsmath, showpacs, showkeys, a4paper, nofootinbib, superscriptaddress, 11pt, raggedbottom]{revtex4-2}

\usepackage{braket}
\usepackage{bm}
\usepackage{graphicx}
\usepackage{slashed}
\usepackage{subfig}
\usepackage{mathrsfs}
\usepackage{color}
\usepackage{mathtools}
\usepackage[normalem]{ulem}
\usepackage{simplewick}

\def\beq{\begin{equation}}
\def\eeq{\end{equation}}
\def\beqa{\begin{eqnarray}}
\def\eeqa{\end{eqnarray}}

\begin{document}

\title{\Large Dilaton-induced open quantum dynamics}

\author{Christian K\"{a}ding}
\email{christian.kaeding@tuwien.ac.at}
\affiliation{Technische Universit\"at Wien, Atominstitut, Stadionallee 2, 1020 Vienna, Austria}

\author{Mario Pitschmann}
\email{mario.pitschmann@tuwien.ac.at}
\affiliation{Technische Universit\"at Wien, Atominstitut, Stadionallee 2, 1020 Vienna, Austria}

\author{Caroline Voith}
\email{e1607478@student.tuwien.ac.at}
\affiliation{Technische Universit\"at Wien, Atominstitut, Stadionallee 2, 1020 Vienna, Austria}

\begin{abstract}
In modern cosmology, scalar fields with screening mechanisms are often used as explanations for phenomena like dark energy or dark matter. Amongst a zoo of models, the environment dependent dilaton, screened by the Polyakov-Damour mechanism, is one of the least constrained ones. Using recently developed path integral tools for directly computing reduced density matrices, we study the open quantum dynamics of a probe, modelled by another real scalar field, induced by interactions with an environment comprising fluctuations of a dilaton. As the leading effect, we extract a correction to the probe's unitary evolution, which can be observed as a frequency shift. Assuming the scalar probe to roughly approximate a cold atom in matter wave interferometry, we show that comparing the predicted frequency shifts in two experimentally distinct setups has the potential to exclude large parts of the dilaton parameter space.

\end{abstract}


\maketitle


\section{Introduction}

In modern cosmology, scalar-tensor theories of gravity \cite{Fujii2003}, in which scalar fields are coupled to the gravitational metric tensor, are often proposed as potential solutions for the problems of dark matter and dark energy, see Refs.\,\,\cite{Clifton2011,Joyce2014} for an overview. In many cases, such scalar-tensor theories lead to a universal coupling of the scalar to the trace  $T^\mu_{\,\,\,\mu}$ of the matter energy-momentum tensor. Consequently, these theories predict the existence of gravity-like fifth forces. However, to date, we did not observe any additional fundamental forces beyond the four known ones, which is reflected in strong Solar System constraints \cite{Dickey1994,Adelberger2003,Kapner2007}.
\\
A phenomenologically interesting way of explaining the apparent absence of fifth forces in our Solar System is via a so-called screening mechanism. Such mechanisms can arise in non-linear scalar field theories, and induce a screening of the scalar's fifth force, i.e.\,\,render it to be very weak, in environments with large values of $T^\mu_{\,\,\,\mu}$, for example, for high densities $\rho^{\text{ext}} = -T^\mu_{\,\,\,\mu}$ in dust-dominated regions like our Solar System. By now, there are several scalar models which exhibit a screening mechanism \cite{BurrageSak,Brax:2021wcv}. Amongst the most prominent so-called screened scalar fields are: the chameleon \cite{Khoury2003,Khoury20032}; the symmetron \cite{Dehnen1992, Gessner1992, Damour1994, Pietroni2005, Olive2008, Brax2010,Hinterbichler2010,Hinterbichler2011}, whose fifth force has recently been studied as an alternative to particle dark matter \cite{Burrage2016_2,OHare:2018ayv,Burrage2018Sym,Kading2023}; the environment dependent dilaton \cite{Damour1994,Gasperini:2001pc,Damour:2002nv,Damour:2002mi,Brax:2010gi,Brax:2011ja,Brax2022}; and the galileon \cite{Dvali2000,Nicolis2008,Ali2012}. Many of these models have already been or are proposed to be tested in a plethora of experiments, see, for example, Refs.\,\,\cite{Burrage:2016bwy,BurrageSak,Pokotilovski:2012xuk,Pokotilovski:2013tma,Burrage:2014oza,Hamilton:2015zga,Lemmel:2015kwa,Burrage:2015lya,Elder:2016yxm,Ivanov:2016rfs,Burrage:2016rkv,Jaffe:2016fsh,Brax:2017hna,Sabulsky:2018jma,Brax:2018iyo,Cronenberg:2018qxf,Hartley2019,Pitschmann:2020ejb}. In addition, in recent years, research programs were initiated to study screened scalars as quantum fields \cite{Brax2018quantch,Burrage2018,Burrage2019,Kading2019}, and it was even suggested to investigate screened scalar-tensor theories in analogue gravity simulations \cite{Hartley2018}. 
\\
Ref.\,\,\cite{Burrage2018} presented the idea for using open quantum dynamical effects \cite{Breuer2002} like frequency shifts or decoherence \cite{Schlosshauer}, which might be induced by interactions with an environment comprising light scalar fluctuations, for further constraining chameleon screened scalar fields in atom interferometry experiments. Besides presenting phenomenological research, Ref.\,\,\cite{Burrage2018} also introduced a novel and powerful theoretical tool, namely a practicable and first principle-based method for deriving quantum master equations for reduced density matrices in scalar quantum field theory. This method relies on the Schwinger-Keldysh formalism \cite{Schwinger,Keldysh} and the Feynman-Vernon influence functional \cite{Feynman}. Based on the work in Ref.\,\,\cite{Burrage2018}, Ref.\,\,\cite{Kading2022x} developed a way of circumventing master equations by directly computing reduced density matrix elements with influence functionals. In addition, Ref.\,\,\cite{Kading2022_2} derived a similar technique for computing total density matrices in interacting quantum field theories.
\\
In this article, we look at the environment dependent dilaton screened scalar field as a yet barely constrained model. Following the approach made in Ref.\,\,\cite{Burrage2018}, we study the open quantum dynamics of a probe real scalar field $\phi$ when interacting with an environment comprising fluctuations $\chi$ of a dilaton field $X$. More precisely, we use the scalar $\phi$ as a proxy for an atom in an atom interferometry experiment (or any other object in a matter interferometer, e.g., neutrons) and, applying the formalism from Ref.\,\,\cite{Kading2022x}, check how its reduced density matrix evolves from an initial time $0$ to a final time $t$ due to the hypothetical interaction with a dilaton. We then read off the most dominant open quantum dynamical effect induced in the probe system $\phi$ in order to get an estimate of the dilaton parameter space that could be constrained experimentally this way. 
\\
Certainly, using a scalar field $\phi$ as a proxy for a more complex object like an atom has its shortcomings. However, since the dilaton only couples to the trace of the atom's energy-momentum tensor, a scalar field is a rough but, in this case, suitable first approximation. In order to improve this approximation, we consider the probe system being restricted to a single-particle subspace and ignore all $\phi$-loop diagrams, which naturally appear in a scalar quantum field theory, but are unrealistic for a complex and stable object like a cold atom. 
\\
Providing a glance at the full potential that open quantum dynamical effects can have for constraining physics beyond the standard models of particles and cosmology, this article presents a first application of Ref.\,\,\cite{Kading2022x}, and motivates future studies with more realistic and sophisticated probe models.   
\\
The article is structured as follows: at first, we review the environment dependent dilaton model in Sec.\,\,\ref{sec:dil}. Subsequently, we take the required formula from Ref.\,\,\cite{Kading2022x}, discuss and evaluate it in Sec.\,\,\ref{sec:open}, and then use the results for estimating constraints for the dilaton parameter space in Sec.\,\,\ref{sec:const}. Finally, in Sec.\,\,\ref{sec:Conclusion}, we draw our conclusions.


\section{Environment dependent dilatons}
\label{sec:dil}

Dilatons originate from string theory \cite{Damour1994,Damour:2002nv,Damour:2002mi}, but have found applications in cosmology \cite{Gasperini:2001pc,Brax:2011ja} and were shown, similar to symmetrons, to be subject to the Polyakov-Damour mechanism \cite{Damour1994}, which coins them to be environment dependent \cite{Brax:2010gi} and therefore a screened scalar field model.
\\
The environment dependent dilaton has an effective potential \cite{Joyce2014}
\begin{eqnarray}
\label{eq:dilpotvar}
    V_{\text{eff}}(\varphi;\rho^{\text{ext}}) &=& \bar{V_0} \, e^{-\varphi/M_{Pl}} + \frac{(\varphi-\varphi_*)^2}{2 \mathcal{M}^2}\rho^{\text{ext}}\,\,\,,
\end{eqnarray} 
where $\bar{V_0}$ is a constant potential with a decreasing exponential function $\exp(-\varphi/M_{Pl})$ as required by the strong coupling limit of string theory. Here, $M_{Pl}$ denotes the Planck mass, $\mathcal{M}$ is a coupling constant with dimension of a mass, for which we require $\varphi \ll \mathcal{M}$, and $\varphi_*$ is a constant field value around which the coupling of the dilaton to matter becomes feeble or even vanishes entirely for $\varphi=\varphi_*$. This decoupling is the essence of the Polyakov-Damour mechanism and leads to a suppression of the dilaton fifth force.
\\
The fact that a larger value of $\rho^{\text{ext}}$ actually corresponds to a weaker coupling to matter can more easily be seen when substituting $\lambda X := \varphi-\varphi_*$ into Eq.\,\,(\ref{eq:dilpotvar}). Introducing  
\begin{eqnarray}
\label{eq:selfint}
V(X) \,:=\, V_0 \,e^{-\lambda X/M_{Pl}}  &:=& 
 \bar{V_0} \, e^{-(\lambda X+\varphi_*)/M_{Pl}}
\end{eqnarray}
with $\lambda$ being a dimensionless coupling constant, we find \cite{Brax2022}
\begin{eqnarray}
\label{eq:dilpotX}
V_{\text{eff}}(X;\rho^{\text{ext}}) 
&=& V_0 \,e^{-\lambda X/M_{Pl}} + \frac{A_2 \rho^{\text{ext}}}{2M_{Pl}^2} X^2\,\,\,,
\end{eqnarray}
where a dimensionless constant  $A_2 := \lambda^2 M_{Pl}^2/\mathcal{M}^2 $ was introduced that has to fulfill $A_2 \gg 1$ in order to circumvent existing Solar System-based constraints on scalar fifth forces \cite{Brax2010}. From Eq.\,\,(\ref{eq:dilpotX}) we find the dilaton vacuum expectation value (vev) as \cite{Brax2022}
\begin{eqnarray} 
\label{eq:dilvev}
\langle X \rangle &=& \frac{M_{Pl}}{\lambda} W\left( \frac{\lambda^2 V_0}{A_2 \rho^{\text{ext}}} \right)
\end{eqnarray} 
with the Lambert $W$-function
\begin{eqnarray} 
W(x) &=& \sum_{n=1}^{\infty} \frac{(-n)^{n-1}}{n!} x^n\,\,\,.
\end{eqnarray} 
Since $X = \langle X \rangle + \chi$, where $\chi$ is a small perturbation, Eq.\,\,(\ref{eq:dilvev}) lets us conclude that $X$ decreases when $\rho^{\text{ext}}$ increases. Consequently, the leading $\langle X \rangle^2$-term in Eq.\,\,(\ref{eq:dilpotX}) becomes smaller for larger values of $\rho^{\text{ext}}$, which indicates a decoupling between dilaton and matter, i.e.\,\,a screening of the fifth force.
\\
As a scalar-tensor theory, the dilaton is coupled to the gravitational metric tensor via 
$\tilde{g}_{\mu \nu} = A^2(X) g_{\mu \nu},
$
where $\tilde{g}$ and $g$ denote the metric in the Jordan and Einstein frame \cite{Fujii2003}, respectively, and the conformal factor is given by 
\begin{eqnarray}
\label{eq:conffact}
    A(X) &=& 1 + \frac{A_2}{2M_{Pl}^2} X^2\,\,\,.
\end{eqnarray}
The total Einstein frame action describing gravity, the scalar $X$ and matter is given by \cite{Khoury2003,Khoury20032}
\begin{eqnarray} 
S &=& \int  d^4x \sqrt{-g} \left[ \frac{1}{2} M_{Pl}^2 R - \frac{1}{2} g^{\mu \nu} \partial_{\mu}X\partial_{\nu}X - V(X) \right]
+\int d^4x \sqrt{-g} A^4(X) \Tilde{\mathcal{L}}_M (\Tilde{\phi}, A^2(X) g_{\mu \nu})
\,\,\,,\,\,\,\,\,\,
\end{eqnarray} 
where the last term describes the dynamics of the Jordan frame matter field $\tilde{\phi}$, which in our case is a real scalar, and its interactions with the dilaton. Here, the potential $V(X)$ is the self-interaction of $X$ given by Eq.\,\,(\ref{eq:selfint}), and for the Jordan frame matter Lagrangian we use
\begin{eqnarray} 
\Tilde{\mathcal{L}}_M &=&-\frac{1}{2} \Tilde{g}^{\mu \nu} \partial_{\mu} \Tilde{\phi} \partial_{\nu} \Tilde{\phi} - \frac{1}{2} \Tilde{M}^2 \Tilde{\phi}^2\,\,\,.
\end{eqnarray} 
Since we will have to work with the dilaton perturbatively if we later want to discuss it in the formalism presented in Ref.\,\,\cite{Kading2022x}, we are required to assume $\lambda X/M_{Pl} \ll 1$, such that
we can expand 
\begin{eqnarray}
\label{eq:potexpand}
    V(X) &\approx& V_0 \left[ 1 -\frac{\lambda }{M_{Pl}}X + \frac{1}{2} \frac{\lambda^2 }{M_{Pl}^2}X^2 + \mathcal{O}\bigg( \frac{\lambda^3 }{M_{Pl}^3}X^3\bigg) \right]\,\,\,.
\end{eqnarray}
Following the general procedure outlined in Ref.\,\,\cite{Burrage2018} for Eqs.\,\,(\ref{eq:conffact}) to (\ref{eq:potexpand}), while only keeping operators of dimension $4$ or lower, we can derive Einstein frame free actions for the matter field $\phi$ and the dilaton fluctuation $\chi$ as well as an interaction action between both types of fields: 
\begin{eqnarray}
S_{\phi}[\phi] &:=& \int_x \left[ - \frac{1}{2} g^{\mu \nu} \partial_{\mu} \phi \partial_{\nu} \phi - \frac{1}{2} M^2 \phi^2 \right]\,\,\,, 
\\
S_{\chi}[\chi] &:=& \int_x \left[ - \frac{1}{2} g^{\mu \nu} \partial_{\mu} \chi \partial_{\nu} \chi - \frac{1}{2} m^2 \chi^2 \right]\,\,\,, 
\\
\label{eq:interact}
S_{\text{int}}[\phi, \chi] &:=& \int_{x \in \Omega_t} \left[ -\frac{1}{2} \alpha_1 M\chi \phi^2 -\frac{1}{4}\alpha_2 \chi^2 \phi^2 \right]\,\,\,, 
\end{eqnarray}
where, using $g_{\mu\nu} \equiv \eta_{\mu\nu}$ from now on,
\begin{eqnarray}
    \int_x &:=& \int d^4x\,\,\,, 
\end{eqnarray}
the masses are 
\begin{eqnarray}
\label{eq:masses}
M^2 &:=& \left( 1+\frac{A_2}{M_{Pl}^2} \langle X \rangle^2 \right) \Tilde{M}^2\,\,\,, 
\,\,\,\,\,\,\,\,\,
m^2 \,\approx\, \frac{1}{M_{Pl}^2} \left( V_0 \lambda^2 + A_2 \rho^{\text{ext}} \right)\,\,\,, 
\end{eqnarray}
the coupling constants take on the forms 
\begin{eqnarray}
\label{eq:alphas}
\alpha_1 &:=& 2M \frac{A_2}{M_{Pl}^2} \langle X \rangle \left( 1-\frac{A_2}{M_{Pl}^2} \langle X \rangle^2 \right)\,\,\,, 
\,\,\,\,\,\,\,\,\,
\alpha_2 \,:=\,  \frac{M}{\langle X \rangle} \alpha_1 \,\,\,,
\end{eqnarray}
and we restrict integrations over spacetime coordinates to\, $\Omega_t := [0,t] \times \mathbb{R}$\, since we are only interested in the finite interval between initial time $0$ and final time $t$. Note that at the order considered in the expansion made in Eq.\,\,(\ref{eq:potexpand}) the dilaton does not self-interact and, therefore, we do not need to consider a self-interaction action for $\chi$.


\section{Open quantum dynamics}
\label{sec:open}

\begin{figure} [htbp]
\centering
    \subfloat[][]{\includegraphics[width=4.5cm]{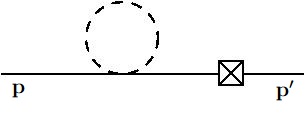}}
    \qquad
    \subfloat[][]{\includegraphics[width=4.5cm]{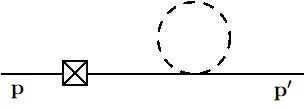}}
  
    \subfloat[][]{\includegraphics[width=4.5cm]{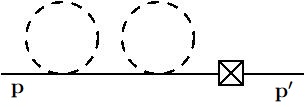}}
    \qquad
    \subfloat[][]{\includegraphics[width=4.5cm]{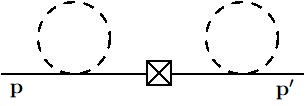}}
    \qquad
    \subfloat[][]{\includegraphics[width=4.5cm]{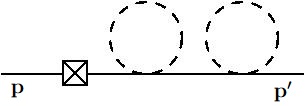}}
\caption{Diagrammatic representation of of the last two terms in the square brackets in Eq.\,\,(\ref{eq:partres}); solid/dashed lines represent $\phi$-/$\chi$-propagators. Crossed boxes depict insertions of the initial density matrix.}
    \label{fig:diag}
\end{figure}

Most realistic quantum systems should be treated as open, i.e.\,\,as interacting with one or several environments. Such interactions can induce open quantum dynamical effects like frequency (or phase) shifts or decoherence in a system. While the theory of open quantum systems naturally finds many applications in non-relativistic quantum physics, see e.g.\,\,Ref.\,\cite{Carmichael,Gardiner2004,Walls2008,Aolita2015,Goold2016,Werner2016,Huber2020}, it also receives increased attention in quantum field theory \cite{Calzetta2008,Koksma2010,Koksma2011,Sieberer2016,Marino2016,Baidya2017,Burrage2018,Nagy2020,Jana2021,Fogedby2022}, and adjacent areas like Early Universe cosmology \cite{Lombardo1,Lombardo2,Lombardo3,Boyanovsky1,Boyanovsky2,Boyanovsky3,Boyanovsky4,Burgess2015,Hollowood,Binder2021,Brahma:2022yxu,Brahma:2021mng,Colas:2022hlq}, black holes \cite{Yu2008,Lombardo2012,Jana2020,Agarwal2020,Kaplanek2020,Burgess2021,Kaplanek2021}, or heavy-ion physics \cite{Brambilla1,Brambilla2,Yao2018,Yao2020,Akamatsu2020,DeJong2020,Yao2021,Brambilla2021,Griend2021,Yao2022}.
\\
Quantum systems, especially open ones, are often described by density operators $\hat{\rho}(t)$, which are advantageous over a wave function description since they can not only describe pure but also mixed states. This is necessary in order to fully capture phenomena like decoherence. In case of an open quantum system, we usually work with reduced density operators, which are obtained by tracing out the environmental degrees of freedom.
\\
Projecting a (reduced) density operator into a basis, for example, a single-particle momentum basis as we will use in this article, gives elements of a density matrix:
\begin{eqnarray}
\rho(\mathbf{p};\mathbf{p}';t)  &=& \bra{\mathbf{p}}\hat{\rho}(t) \ket{\mathbf{p}'}\,\,\,.
\end{eqnarray}
The time evolution of density matrices can be described by quantum master equations. However, such equations are often analytically intricate or even impossible to solve. In order to avoid such complications, using the technology presented in Ref.\,\,\cite{Burrage2018}, Ref.\,\,\cite{Kading2022x} developed a formalism that enables us to directly compute reduced density matrices in terms of the Feynman-Vernon influence functional \cite{Feynman}, which itself is based on the Schwinger-Keldysh formalism \cite{Schwinger,Keldysh}.
\\
In this article, we consider an open quantum system $\phi$, which is a proxy for an atom (or another, compared to the dilaton relatively heavy matter particle like a neutron) with zero temperature in order to justify a restriction to the single-particle subspace. As was pointed out in Refs.\,\,\cite{Burrage2018} and \cite{Kading2022x}, in this case, contractions of the system field can only give rise to Feynman and Dyson propagators: 
\begin{eqnarray}
\contraction{}{\phi}{^+_x}{\phi}\phi^+_x\phi^+_y
&=& 
D^{++}_{xy} \,=\, D^\mathrm{F}_{xy} \,=\, - \mathrm{i}\int_k \frac{e^{\mathrm{i}k\cdot (x-y)}}{k^2+M^2-\mathrm{i}\epsilon}\,\,\,,
\\
\contraction{}{\phi}{^-_x}{\phi}\phi^-_x\phi^-_y
&=& 
D^{--}_{xy} \,=\, D^\mathrm{D}_{xy} \,=\, + \mathrm{i}\int_k \frac{e^{\mathrm{i}k\cdot (x-y)}}{k^2+M^2+\mathrm{i}\epsilon}
\,\,\,,
\end{eqnarray}
where 
\begin{eqnarray}
    \int_{k} &:=&\int\frac{d^4 k}{(2\pi)^4}\,\,\,,
\end{eqnarray}
and $+$ and $-$ denote the positive and negative branches of the Schwinger-Keldysh closed time path, respectively. 
\\
The system is interacting with an environment comprised of dilaton fluctuations $\chi$. As was also done in Ref.\,\,\cite{Burrage2018} for the chameleon, we assume the dilaton to have a, in general, non-zero temperature $T$, for example, due to thermalization with the walls of a vacuum chamber in an atom interferometry experiment. This means, open system and environment are out of thermal equilibrium in our discussion. Since environmental degrees of freedom can be contracted for any combination of $+$ and $-$ \cite{Burrage2018,Kading2022x}, the dilaton can give Feynman and Dyson, but also Wightman propagators: 
\begin{eqnarray}\label{eq:FeynmanProp}
\contraction{}{\chi}{^+_x}{\chi}\chi^+_x\chi^+_y
&=&  \Delta^{++}_{xy}  \,=\, \Delta^{\rm F}_{xy}
\,=\, -\mathrm{i} \int_k e^{ik\cdot (x-y)}\left[\frac{1}{k^2+m^2-\mathrm{i}\epsilon} +2\pi \mathrm{i} f(|k^0|) \delta(k^2+m^2) \right]\,\,\,,
\\
\contraction{}{\chi}{^-_x}{\chi}\chi^-_x\chi^-_y
 &=&  \Delta^{--}_{xy} \,=\, \Delta^{\rm D}_{xy}
\,=\, + \mathrm{i} \int_k e^{\mathrm{i}k\cdot (x-y)}\left[\frac{1}{k^2+m^2+\mathrm{i}\epsilon} -2\pi \mathrm{i} f(|k^0|) \delta(k^2+m^2) \right]\,\,\,,
\\
\contraction{}{\chi}{^+_x}{\chi}\chi^+_x\chi^-_y
&=& \langle\chi_{y} \chi_{x}\rangle \,=\,  \Delta^{+-}_{xy} \,=\, \Delta^<_{xy}
\,=\, \int_k e^{\mathrm{i}k\cdot (x-y)} 2\pi \text{sgn}(k^0)f(k^0) \delta(k^2+m^2)\,\,\,,
\\
\label{eq:WMProp}
\contraction{}{\chi}{^-_x}{\chi}\chi^-_x\chi^+_y
&=&  \langle\chi_{x} \chi_{y}\rangle  \,=\, \Delta^{-+}_{xy} \,=\, \Delta^>_{xy}~=~ \Delta^<_{yx} \,=\, (\Delta^<)^*_{xy}\,\,\,,
\end{eqnarray}
where 
\begin{eqnarray}
    f(k^0)&:=&\frac{1}{e^{\beta k^0}-1} 
    \,=\,-[1+f(-k^0)]
\end{eqnarray}
is the Bose-Einstein distribution function with $\beta = 1/T$ being the inverse temperature. 
\\
Following Ref.\,\,\cite{Kading2022x}, the reduced density matrix element for $\phi$ at time $t$ in the single-particle momentum subspace, under the assumption that the system's particle number does not change in the interval $[0,t]$, can be obtained by evaluating  
\begin{eqnarray}\label{eq:11DensGen}
\rho(\mathbf{p};\mathbf{p}' ;t)
&=& 
\lim_{\substack{x^{0(\prime)}\,\to\, t^{+}\\y^{0(\prime)}\,\to\, 0^-}}
\int_{\mathbf{k}\mathbf{k}'} \frac{\rho(\mathbf{k};\mathbf{k}';0) }{(2 E_{\mathbf{k}}^\phi)(2 E_{\mathbf{k}'}^\phi)}
\nonumber
\\
&&
\times 
\int_{\mathbf{x}\mathbf{x}'\mathbf{y}\mathbf{y}'} e^{-\mathrm{i}(\mathbf{p}\cdot\mathbf{x}-\mathbf{p}' \cdot\mathbf{x}')+\mathrm{i}(\mathbf{k}\cdot\mathbf{y}-\mathbf{k}'\cdot\mathbf{y}')}
\partial_{x^0,E^\phi_{\mathbf{p}}} \partial_{x^{0\prime},E^\phi_{\mathbf{p}'}}^*\partial_{y^0,E^\phi_{\mathbf{k}}}^*\partial_{y^{0\prime},E^\phi_{\mathbf{k}'}}
\nonumber
\\
&&
\times 
\int\mathcal{D}\phi^{\pm} e^{\mathrm{i}\left\{S_{\phi}[\phi^+]-S_{\phi}[\phi^-]\right\}}\phi^+_x\phi^-_{x'}\mathcal{F}[\phi^\pm;t]\phi^{+}_y\phi^{-}_{y'}
\,\,\,,
\end{eqnarray}
where the Feynman-Vernon influence functional
\begin{eqnarray}\label{eqn:IFaction}
\mathcal{F}[\phi^\pm;t] &=& \left\langle e^{\mathrm{i}\big\{   S_\text{int}[\phi^+,\chi^+;t]- S_\text{int}[\phi^-,\chi^-;t] \big\}} \right\rangle_\chi 
\end{eqnarray}
is given in terms of the expectation value
\begin{eqnarray} 
\langle A[\chi^{a}]\rangle_\chi &:=& \int d\chi^{\pm}_t d\chi^{\pm}_0 \delta(\chi_t^+-\chi_t^-)\rho_\chi [\chi^{\pm}_0;0]
\int^{\chi^{\pm}_t}_{\chi^{\pm}_0} \mathcal{D}\chi^{\pm} A[\chi^{a}]e^{\mathrm{i}\left\{S_{\chi}[\chi^+]-S_{\chi}[\chi^-]\right\}}
\end{eqnarray} 
with $\rho_\chi [\chi^{\pm}_0;0]$ denoting the initial environmental density matrix under the assumption that system and environment were initially uncorrelated. The index of a field, e.g.\,\,$t$ on $\chi_t$, labels the time slice on which the field eigenstate was taken. 
\\
We expand the Feynman-Vernon influence functional up to second order in the coupling constants $\alpha_1$ and $\alpha_2$, and find 
\begin{eqnarray} 
\label{eq:FVIfex}
\mathcal{F}[\phi^\pm;t] &=& 1 - \mathrm{i} \frac{\alpha_2}{4} \sum_{a=\pm} a \int_x (\phi^a_x)^2 \Delta^{\rm F}_{xx} - \frac{1}{2} \sum_{a,b=\pm} ab \int_{xy} \biggl[ \frac{\alpha_1^2 M^2}{4} (\phi^a_x)^2 (\phi^b_y)^2 \Delta^{ab}_{xy} 
\nonumber
\\
&&
+ \frac{\alpha_2^2}{16} (\phi^a_x)^2 (\phi^b_y)^2 \left( \Delta^{\rm F}_{xx} \Delta^{\rm F}_{yy} +2(\Delta^{ab}_{xy})^2 \right) \bigg] + \mathcal{O}\left(\alpha^3_{1,2}\right)\,\,\,.
\end{eqnarray} 
Substituting Eq.\,\,(\ref{eq:FVIfex}) into Eq.\,\,(\ref{eq:11DensGen}), using Wick's theorem \cite{Wick}, and dropping every term containing $\phi$-loops, we obtain 
\begin{eqnarray} 
\label{eq:rhoprop}
\rho(\mathbf{p}; \mathbf{p}';t) &=&
\lim_{\substack{x^{0(\prime)}\to t^+\\ y^{0(\prime)} \to 0^-}} \int_{\mathbf{k}\mathbf{k}'} \frac{\rho(\mathbf{k};\mathbf{k}';0) }{(2 E_{\mathbf{k}}^\phi)(2 E_{\mathbf{k}'}^\phi)} 
\nonumber
\\
&&
\times \int_{\mathbf{x}\mathbf{x}'\mathbf{y}\mathbf{y}'} e^{-\mathrm{i} (\mathbf{p}\cdot\mathbf{x}-\mathbf{p}'\cdot\mathbf{x}')+\mathrm{i} (\mathbf{k}\cdot\mathbf{y}-\mathbf{k}'\cdot\mathbf{y}')} \partial_{x^0,E_\mathbf{p}^\phi} \partial^*_{x^{0\prime},E_{\mathbf{p}'}^\phi} \partial^*_{y^0,E_\mathbf{k}^\phi} \partial_{y^{0\prime},E_{\mathbf{k}'}^\phi} 
\nonumber
\\
&&
\times 
\Bigg\{ D^\mathrm{F}_{xy}D^\mathrm{D}_{x'y'} - \mathrm{i}  \frac{\alpha_2}{2} \int_z \left[ D^\mathrm{F}_{xz}D^\mathrm{F}_{zy}D^\mathrm{D}_{x'y'} - (x,y \longleftrightarrow x',y')^\ast \right] \Delta^{\rm F}_{zz}  
\nonumber
\\
&&
- \frac{\alpha_1^2 M^2}{8} \int_{zz'} \bigg[ 
\Big(
D^\mathrm{D}_{x'y'}\big(8 D^\mathrm{F}_{xz} D^\mathrm{F}_{zz'} D^\mathrm{F}_{z'y} + 2 D^\mathrm{F}_{xy} D^\mathrm{F}_{zz'} D^\mathrm{F}_{zz'} 
\big) \Delta^{\rm F}_{zz'}
+(x,y \longleftrightarrow x',y')^\ast \Big)
\nonumber
\\
&&
\,\,\,\,\,\,\,\,\,\,\,\,\,\,\,\,\,\,\,\,\,\,\,\,\,\,\,\,\,\,\,\,\,\,\,\,
- \big(8 D^\mathrm{F}_{xz} D^\mathrm{F}_{zy} D^\mathrm{D}_{x'z'} D^\mathrm{D}_{z'y'} + 4 D^\mathrm{F}_{xy} D^\mathrm{F}_{zz} D^\mathrm{D}_{x'z'} D^\mathrm{D}_{z'y'} 
\big) \Delta^{+-}_{zz'} 
\bigg] 
\nonumber
\\
&&
- \frac{\alpha_2^2}{32} \int_{zz'} \bigg[ 
\Big( 
D^\mathrm{D}_{x'y'}\big(8 D^\mathrm{F}_{xz} D^\mathrm{F}_{zz'} D^\mathrm{F}_{z'y} + 2 D^\mathrm{F}_{xy} D^\mathrm{F}_{zz'} D^\mathrm{F}_{zz'} 
\big) \big(\Delta^{\rm F}_{zz}\Delta^{\rm F}_{z'z'} + 2(\Delta^{\rm F}_{zz'})^2 \big) 
\nonumber
\\
&&
\,\,\,\,\,\,\,\,\,\,\,\,\,\,\,\,\,\,\,\,\,\,\,\,\,\,\,\,\,\,
+(x,y \longleftrightarrow x',y')^\ast \Big)
\nonumber
\\
&&
\,\,\,\,\,\,\,\,\,\,\,\,\,\,\,\,\,\,\,\,\,\,\,\,\,\,
- \big(8 D^\mathrm{F}_{xz} D^\mathrm{F}_{zy} D^\mathrm{D}_{x'z'} D^\mathrm{D}_{z'y'} + 4 D^\mathrm{F}_{xy} D^\mathrm{F}_{zz} D^\mathrm{D}_{x'z'} D^\mathrm{D}_{z'y'} 
\big) \big(\Delta^{\rm F}_{zz}\Delta^{\rm F}_{z'z'} + 2(\Delta^{+-}_{zz'})^2 \big) 
\bigg]
\Bigg\}\,\,\,.\,\,\,\,\,\,\,\,\,\,\,\,\,\,\,
\end{eqnarray} 
Next, we evaluate the integrals in Eq.\,\,(\ref{eq:rhoprop}). The full result of this computation can be found in Appendix \ref{app:Dens}. However, for the present discussion we focus only on the following terms:
\begin{eqnarray}
\label{eq:partres}
\rho(\mathbf{p}; \mathbf{p}';t) &=&  
\rho(\mathbf{p}; \mathbf{p'};0) e^{-\mathrm{i}(E_{\mathbf{p}}^{\phi} - E_{\mathbf{p'}}^{\phi}) t}
\nonumber
\\
&&
\times
\Bigg[
1 - \frac{\mathrm{i} \alpha_2}{4} \left( \frac{1}{E_{\mathbf{p}}^{\phi}} - \frac{1}{E_{\mathbf{p'}}^{\phi}} \right) t \Delta^\mathrm{F}_{zz} 
- \frac{\alpha_2^2}{32} \left( \frac{1}{E_{\mathbf{p}}^{\phi}} - \frac{1}{E_{\mathbf{p'}}^{\phi}} \right)^2t^2 
\Delta^\mathrm{F}_{zz} \Delta^\mathrm{F}_{z'z'} \Bigg] 
+ \mathcal{O}\left(\alpha^2_{1,2}\right)\,\,\,.\,\,\,\,\,\,
\end{eqnarray}
A diagrammatic depiction of the last two terms in the square brackets can be found in Fig.\,\,\ref{fig:diag}. 
Following Ref.\,\,\cite{Burrage2018}, we deal with the tadpoles in Eq.\,\,(\ref{eq:partres}) by adding a counter term
\begin{eqnarray}
\delta S_{\text{int}}[\phi, \chi] &:=&    \ \frac{\alpha_2}{4} \int_{x}  \Delta^{F(T=0)}_{xx}  \phi^2  
\end{eqnarray}
to the interaction action in Eq.\,\,(\ref{eq:interact}), where $(T=0)$ indicates the temperature-independent part of the $\chi$-propagator. The thermal part 
\begin{eqnarray}
\Delta^{\mathrm{F}(T\neq 0)}_{xx} 
&=&
\int_k 2\pi f(|k^0|) \delta(k^2+m^2)  \,=\, \frac{T^2}{2\pi^2} \int_{m/T}^\infty d\xi \frac{\sqrt{\xi^2-(\frac{m}{T})^2}}{e^\xi -1}
\end{eqnarray}
is by default finite in the ultraviolet. Using this, and identifying the terms in the square brackets of Eq.\,\,(\ref{eq:partres}) as the expansion of an exponential function, we find 
\begin{eqnarray}
\label{eq:resat}
 \rho(\mathbf{p}; \mathbf{p}';t) &=&  
\rho(\mathbf{p}; \mathbf{p'};0) \exp\left\{-\mathrm{i}\Bigg[E_{\mathbf{p}}^{\phi} - E_{\mathbf{p'}}^{\phi} 
+
\frac{\alpha_2}{4} \left( \frac{1}{E_{\mathbf{p}}^{\phi}} - \frac{1}{E_{\mathbf{p'}}^{\phi}} \right) \Delta^{\mathrm{F}(T\neq 0)}_{zz}\Bigg] t\right\}  
+ \mathcal{O}\left(\alpha^2_{1,2}\right)
\,\,\,.
\end{eqnarray}
Now we recall that we performed all computations with the rescaled mass $M$ as given in Eq.\,\,(\ref{eq:masses}). However, cp.\,\,with Ref.\,\,\cite{Burrage2018}, experiments are actually sensitive to the absolute mass $\tilde{M}$. Rewriting Eq.\,\,(\ref{eq:resat}), we obtain 
\begin{eqnarray}
\label{eq:resed}
 \rho(\mathbf{p}; \mathbf{p}';t) &=&  
\rho(\mathbf{p}; \mathbf{p'};0) \exp\left\{-\mathrm{i}\Bigg[\tilde{E}_\mathbf{p}^\phi - \tilde{E}_{\mathbf{p}'}^\phi
\right.
\nonumber
\\
&&
\,\,\,\,\,\,\,\,\,\,\,\,\,\,\,\,\,\,\,\,\,\,\,\,\,\,\,\,\,\,\,\,\,\,\,\,\,\,\,\,\,\,\,\,\,\,\,\,
\left.
+ 
\tilde{M}^2 \frac{A_2}{2M_{Pl}^2}
\left(  \langle X\rangle^2 
+
\Delta^{\mathrm{F}(T\neq 0)}_{zz} \right)
 \left( \frac{1}{\tilde{E}_\mathbf{p}^\phi} - \frac{1}{\tilde{E}_{\mathbf{p}'}^\phi} \right)
\Bigg] t\right\}  
+ \mathcal{O}\left(\alpha^2_{1,2}\right)
\,\,\,,\,\,\,\,\,\,\,\,\,\,\,\,
\end{eqnarray}
where $\tilde{E}_\mathbf{p}^\phi=\sqrt{\mathbf{p}^2+\tilde{M}^2}$, we replaced $\alpha_2$ by Eq.\,\,(\ref{eq:alphas}), and neglected the term $\sim \mathcal{O}(A_2^2\tilde{M}^2 \langle X \rangle^2/M_{Pl}^4)$. What we found in Eq.\,\,(\ref{eq:resed}) is a first order correction to the unitary evolution term, i.e.\,\,a phase shift $\Delta u \cdot t$ with a frequency shift $\Delta u$. Other effects like de-/recoherence and momentum diffusion first appear at second order, see Appendix \ref{app:Dens}, and are therefore expected to be subdominant.


\section{Predicted constraints}
\label{sec:const}

\begin{figure} [htbp]
\centering
    \subfloat[][]{\includegraphics[width=7.8cm]{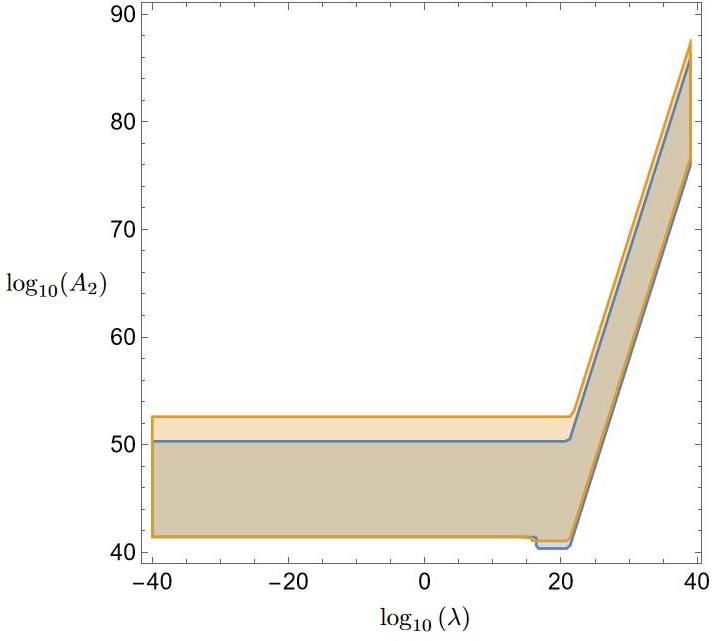}}
    \qquad
    \subfloat[][]{\includegraphics[width=7.8cm]{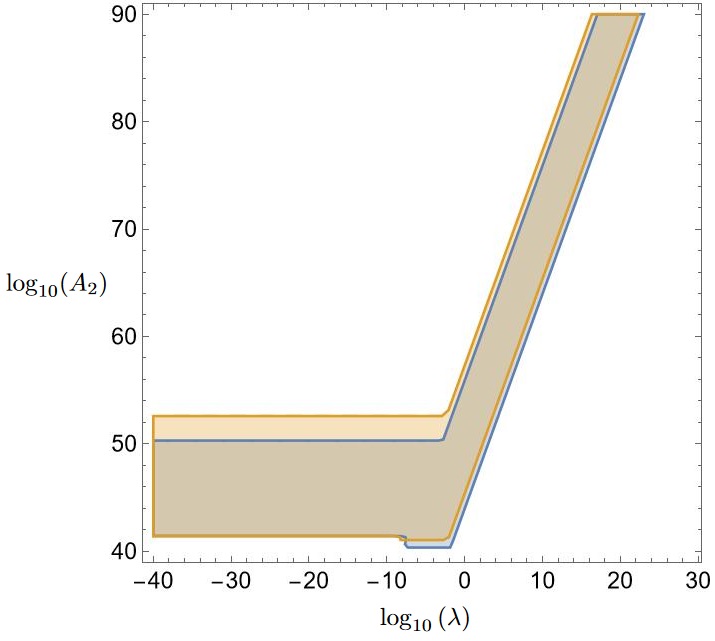}}
\caption{Exclusion plots for the environment dependent dilaton $(\lambda,A_2)$-parameter space resulting from differences in frequency shifts $u(P_1,T_2;P_1,T_1)$ (blue) and $u(P_1,T_4;P_1,T_1)$ (orange); (a):\,\,\,$V_0 = 1\,\text{eV}^4$, (b): $V_0 = 1\,(\text{MeV})^4$}
    \label{fig:lowpress}
\end{figure}

\begin{figure} [htbp]
\centering
    \subfloat[][]{\includegraphics[width=7.8cm]{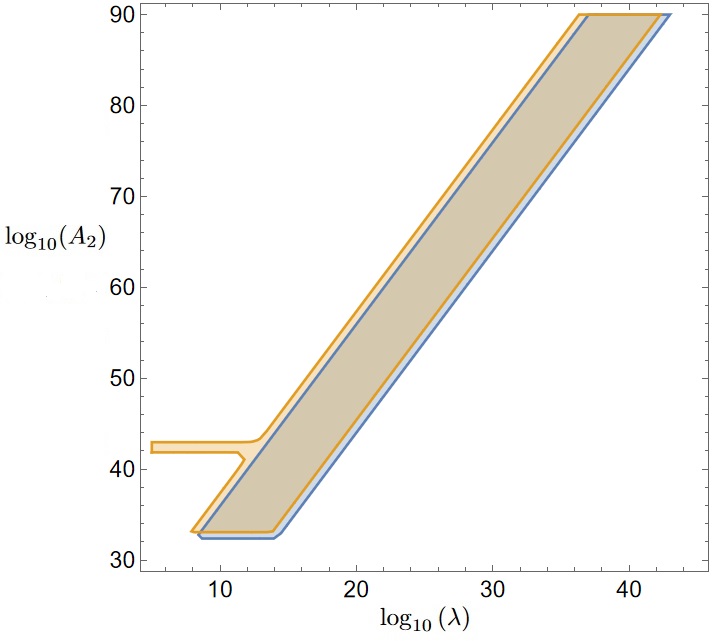}}
    \qquad
    \subfloat[][]{\includegraphics[width=7.8cm]{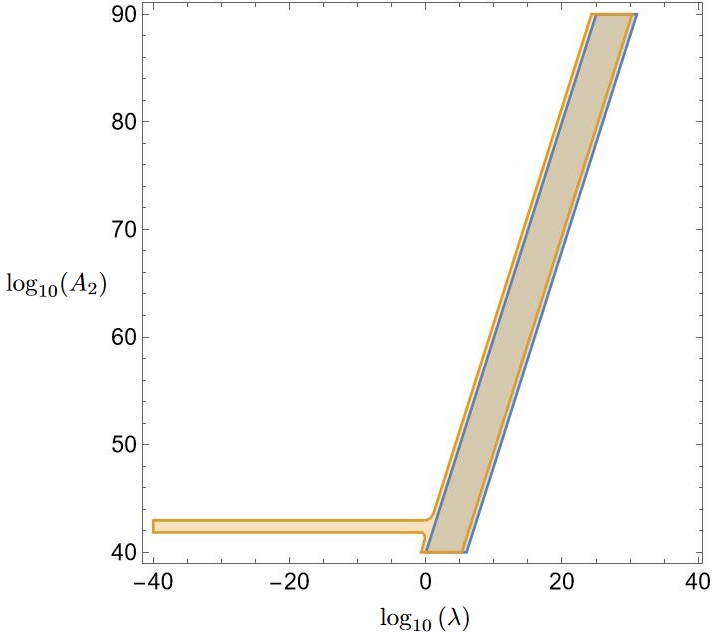}}
\caption{Exclusion plots for the environment dependent dilaton $(\lambda,A_2)$-parameter space resulting from differences in frequency shifts $u(P_2,T_2;P_2,T_1)$ (blue) and $u(P_2,T_4;P_2,T_1)$ (orange); (a):\,\,\,$V_0 = 1\,(\text{keV})^4$, (b): $V_0 = 1\,(\text{MeV})^4$}
    \label{fig:medpress}
\end{figure}

\begin{figure} [htbp]
\centering
    \includegraphics[width=7.8cm]{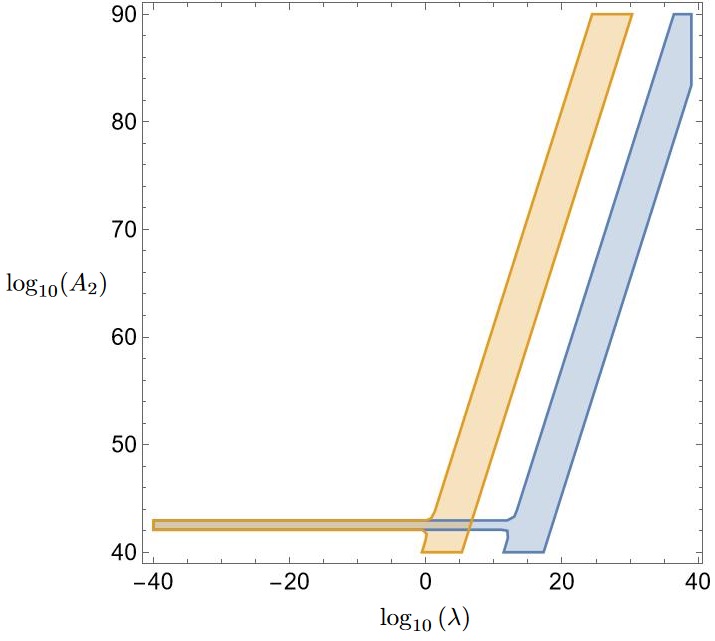}
\caption{Exclusion plots for the environment dependent dilaton $(\lambda,A_2)$-parameter space resulting from difference in frequency shifts $u(P_2,T_4;P_2,T_3)$; the blue plot represents $V_0 = 1\,\text{eV}^4$ and the orange plot $V_0 = 1\,(\text{MeV})^4$.}
\label{fig:hight}
\end{figure}

Atom interferometry experiments have successfully been used to constrain a variety of screened scalar field models \cite{Jaffe:2016fsh,Elder:2016yxm,Burrage:2015lya,Burrage:2016rkv,Hamilton:2015zga,Burrage:2014oza,Sabulsky:2018jma}. As was also done in Ref.\,\,\cite{Burrage2018}, we now estimate whether the frequency shift found in Eq.\,\,(\ref{eq:resed}) could potentially be observed in an atom interferometer. Since such experiments use low-energy atoms, we make a non-relativistic approximation $\tilde{M}^2 \gg \mathbf{p}^2$, such that
\begin{eqnarray}
 \frac{1}{\tilde{E}_\mathbf{p}^\phi}  &\approx&
 \frac{1}{\tilde{M}} \left( 1- \frac{\mathbf{p}^2}{2\tilde{M}^2}\right)\,\,\,,\,\,\,\,\,\,\,\,\,\frac{1}{\tilde{E}_\mathbf{p}^\phi} - \frac{1}{\tilde{E}_{\mathbf{p}'}^\phi} \,\approx\,\frac{v^2}{2\Tilde{M}} 
\end{eqnarray}
with $v := \frac{||\mathbf{p}|-|\mathbf{p}'||}{\Tilde{M}}$ being the speed difference between the two atomic states. Substituting this into Eq.\,\,(\ref{eq:resed}), gives for the frequency shift:
\begin{eqnarray}
\label{eq:shift}
\Delta u 
&:=& 
 \tilde{M} \frac{A_2}{4M_{Pl}^2} \left[ \langle X\rangle^2 + \Delta^{\mathrm{F}(T\neq0)}_{zz} \right]v^2 \,\,\,.
\end{eqnarray}
For the estimation we choose commonly used values for the parameters of the experiment: a vacuum chamber with radius $L = 10$ cm \cite{Burrage:2015lya,Burrage:2014oza}, and a Rubidium-87 atom with a mass of $\tilde{M}=87\, m_u$ \cite{Sabulsky:2018jma}, where $m_u$ is the atomic mass unit, as the probe object. For the speed difference between the atomic states we use $v=50$ mm $\text{s}^{-1}$ \cite{bcca-m22}. In addition, from Refs.\,\,\cite{bcca-m22,eymkl15} we infer that in modern atom interferometry experiments frequency shifts as small as $\Delta u_\text{min} \approx 10^{-8}$ Hz can be measured.
For the dilaton's vev we follow the ansatz made in Ref.\,\,\cite{Burrage2018} and assume that the experiment takes place in a sufficiently small part in the center of the vacuum chamber, such that we can assume $\langle X\rangle$ to be approximately spatially constant. In addition, we only consider parts of the parameter space for which the Compton wavelength $\lambda_\mathrm{C} = 1/m$ of the dilaton fulfils $\lambda_\mathrm{C} \leq L$, such that the field is able to take on its proper vev and not an intermediate value. In order to also consider cases for which $\lambda_\mathrm{C} > L$, we would need a sufficiently good model for the evolution of the field profile from within the vacuum chamber walls to the center of the chamber, as was developed for the chameleon, for example, in Ref.\,\,\cite{Elder:2016yxm}. However, the sophisticated numerical studies required for this are beyond the scope of the present article.
\\
A frequency shift is only a useful observable if we can compare it to another measurement without any frequency shift or at least with a frequency shifted by a different magnitude. Since the expression in Eq.\,\,(\ref{eq:shift}) depends on some parameters that can easily be varied, two measurements with different predicted frequency shifts can be compared. If the resulting predicted difference in frequency shifts cannot be observed, this puts constraints on the dilaton parameter space. The parameters that we choose to vary are the temperature $T$ of the vacuum chamber walls, which affects the thermal part of the $\chi$-propagators, and the pressure $P$, which, if we assume that the residual gas inside the vacuum chamber has thermalized with the walls, determines the value of $\rho^\text{ext}$ due to $\rho^\text{ext} = P m_\text{mol}/(T R_\text{gas})$ with $m_\text{mol}$ being the molar mass of the residual gas and $R_\text{gas}$ the universal gas constant, and consequently leads to different values of $\langle X\rangle$ and $m$. For the temperature, we consider four different values: $T_1 = 0.5 \times 10^{-3}$ K, $T_2 = 100$ K, $T_3 = 300$ K, and $T_4 = 500$ K. Ref.\,\,\cite{Sabulsky:2018jma} reports a residual $\text{H}_2$ ($m_\text{mol}(\text{H}_2) = 2.016$ g/mol \cite{hydrogen}) gas pressure of $P_2 = 9.6 \times 10^{-10}$ mbar for their experiment. However, as most extreme values, vacua with $P_1 \approx 10^{-17}$ mbar \cite{lowdens} can be reached, and atom interferometry experiments have been performed in warm vapors of rubidium ($m_\text{mol}(\text{Rb}) = 85.468$ g/mol \cite{rubidium}) atoms with pressures as high as $P_3 \approx 10^{-2}$ mbar \cite{warmvap}. 
\\
We select two pairs of pressure and temperature, and compare the resulting difference in frequency shifts, which can only be observed if
\begin{eqnarray}
\label{eq:psdiff}
|\Delta u(P_a,T_b) - \Delta u(P_c,T_d)| &=:& u(P_a,T_b;P_c,T_d) \,\geq\, \Delta u_\text{min}
\end{eqnarray}
is fulfilled. Furthermore, our previous assumptions require us to demand $\lambda \langle X\rangle/M_{Pl} \ll 1$ and $\sqrt{A_2} \langle X\rangle/M_{Pl} \ll 1$. Using these restrictions, and choosing values for $V_0$, we plot the parts of the $(\lambda,A_2)$-space for which Eq.\,\,(\ref{eq:psdiff}) is fulfilled. The resulting plots for a few selected examples can be found in Figs.\,\,\ref{fig:lowpress} - \ref{fig:hight}. In case of an experimental null result, the shaded areas in these figures exclude the respective parts of  the environment dependent dilaton parameter space. 
\\
In Fig.\,\,\ref{fig:lowpress} we present cases for which the lowest possible pressure $P_1$ is used, but the temperature is varied. For Fig.\,\,\ref{fig:lowpress}(a) we select $V_0 = 1\,\text{eV}^4$, and consider $u(P_1,T_2;P_1,T_1)$ (blue plot) and $u(P_1,T_4;P_1,T_1)$ (orange plot), while for Fig.\,\,\ref{fig:lowpress}(b) we use $V_0 = 1\,(\text{MeV})^4$, and keep everything else the same. Looking at these figures, we observe that in case of a null result, large parts of the dilaton parameter space could be excluded. Comparing frequency shifts with larger temperature differences, i.e.\,\,$(T_4,T_1)$, extends the exclusion area. However, considering the lower temperature difference $(T_2,T_1)$ adds comparatively small areas to the exclusion plots, which cannot be constrained with only $(T_4,T_1)$.   For the $V_0 = 1\,\text{eV}^4$ case, the horizontal exclusion area, which is more than $10$ orders of magnitude in $A_2$ wide, stretches beyond $\lambda = 10^{20}$, while for $V_0 = 1\,(\text{MeV})^4$ it does not even reach $\lambda \approx 1$. 
\\
In Fig.\,\,\ref{fig:medpress} we consider $P_2$ and again vary the temperature, such that blue plots represent $u(P_2,T_2;P_2,T_1)$ and orange ones $u(P_2,T_4;P_2,T_1)$. Fig.\,\,\ref{fig:medpress}(a) depicts the case $V_0 = 1\,(\text{keV})^4$ and (b) $V_0 = 1\,(\text{MeV})^4$. Interestingly, the main difference between blue and orange shaded areas is that each orange one has a narrow (in $A_2$) horizontal section that cannot be found for the blue ones. Similarly to Fig.\,\,\ref{fig:lowpress}, going to higher values of $V_0$ lowers the values of $\lambda$ that can be excluded by this section. However, here, the horizontal stripe is much shorter in Fig.\,\,\ref{fig:medpress}(a) than in (b). 
\\
Finally, in Fig.\,\,\ref{fig:hight} we again consider the pressure $P_2$, but compare frequency shifts for the highest temperatures $T_3$ and $T_4$, i.e.\,\,we look at $u(P_2,T_4;P_2,T_3)$. The blue plot depicts the case $V_0 = 1\,\text{eV}^4$ and the orange one $V_0 = 1\,(\text{MeV})^4$. Both exclusion areas appear to be similar but shifted by more than $10$ orders of magnitude on the $\lambda$-axis.
\\
We do not find any exclusion areas for the highest pressure $P_3$ in combination with the other parameter values we chose. This can be explained by the rather large $\rho^\text{ext}$ that results from the high pressure and temperatures in the considered warm rubidium vapors. A too dense environment suppresses the dilaton field, affecting the non-thermal part in Eq.\,\,(\ref{eq:shift}), and increases its mass, which decreases the thermal part, such that the frequency shift predicted becomes too small to be observable.

\section{Conclusions}
\label{sec:Conclusion}
Screened scalar fields are frequently used models in cosmology. Their screening mechanisms enable them to circumvent tight Solar-System constraints on their gravity-like fifth forces. However, it is possible to probe their parameter spaces in laboratory-based experiments like atom interferometry. While other screened scalar field models like the chameleon or the symmetron are already experimentally well-studied, the parameter space of the environment dependent dilaton, motivated by string theory and screened by the Polyakov-Damour mechanism, is still largely unconstrained.
\\
Recently, Ref.\,\,\cite{Kading2022x} presented a new path integral-based method for directly computing reduced density matrices of open quantum systems. In the present article, we applied this formalism to an open system modelled by the real, massive scalar field $\phi$ coupled to an environment comprising fluctuations $\chi$ of the dilaton field $X$. Following the study in Ref.\,\,\cite{Burrage2018}, it was our intention to use $\phi$ as proxy for an atom in matter wave interferometry. For an atom, this was a rough, but, in this particular physical situation, suitable, first approximation since the dilaton only couples to the mass density. We improved the approximation by restricting the discussion to the single-particle momentum subspace, and neglecting all diagrams containing $\phi$-loops since those would not appear at the considered perturbative order for complex composite objects like atoms. Using Ref.\,\,\cite{Kading2022x}, we computed the single-particle density matrix element in a momentum basis at final time $t$ under the assumption of an initial single atom at time $0$. The resulting expression was presented in Appendix \ref{app:Dens}. However, for the actual discussion, we focused only on the leading effect, which turned out to be a correction to the unitary evolution of $\phi$. In an interferometry experiment, such a correction would be visible as a frequency shift if the measurement was compared to another one with a different or without any shifted frequency. 
\\
For realistic parameters of an atom interferometry experiment, we used this frequency shift in order to predict exclusion plots for the dilaton parameter space. While these predictions should be taken with care due to us approximating an atom by a scalar field, our investigation offers a glimpse at the full potential that open quantum dynamical effects have for studying and constraining physics beyond the standard models of particles and cosmology. Since the predicted exclusion plots look very promising, the present article, besides being the very first practical application of the formalism developed in Ref.\,\,\cite{Kading2022x}, serves as a strong motivation for future studies with more realistic probe systems. Those will provide us with new, powerful tools for our search after screened scalar fields and even other candidate models for dark energy or dark matter.


\begin{acknowledgments}
The authors are grateful to H.~Abele and P.~Haslinger for useful discussions.
This article was supported by the Austrian Science Fund (FWF): P 34240-N, and is based upon work from COST Action COSMIC WISPers CA21106,
supported by COST (European Cooperation in Science and Technology).
\end{acknowledgments}


\appendix

\section{Density matrix elements}
\label{app:Dens}

Evaluating Eq.\,\,(\ref{eq:rhoprop}) gives us
\begin{eqnarray}
&&\rho(\mathbf{p}; \mathbf{p'};t) \,=\, 
\rho(\mathbf{p}; \mathbf{p'};0) e^{-\mathrm{i}(E_{\mathbf{p}}^{\phi} - E_{\mathbf{p'}}^{\phi}) t} \Biggl\{ 1- (2 \pi)^3 \delta^{(3)}(\mathbf{0}) \Biggl[ \frac{\alpha^2_2}{8} \int_{\mathbf{q}} \frac{\sin^2(E_{\mathbf{q}}^{\phi} t)}{4 (E_{\mathbf{q}}^{\phi})^4}  \Delta^\mathrm{F}_{zz} \Delta^\mathrm{F}_{z'z'} 
\nonumber
\\ 
&&
+ \frac{\alpha_1^2 M^2}{2} \int_{\mathbf{k} \mathbf{q}} \frac{1}{2E_{\mathbf{k}}^{\phi} E_{\mathbf{q}}^{\phi} E_{\mathbf{q}-\mathbf{k}}^{\chi}} \Biggl( \frac{\sin^2\left(\frac{(E_{\mathbf{k}}^{\phi} + E_{\mathbf{q}}^{\phi} + E_{\mathbf{q}-\mathbf{k}}^{\chi}) t}{2}\right)}{(E_{\mathbf{k}}^{\phi} + E_{\mathbf{q}}^{\phi} + E_{\mathbf{q}-\mathbf{k}}^{\chi})^2} + \sum_{a=\pm} 
\frac{\sin^2\left(\frac{(E_{\mathbf{k}}^{\phi} + E_{\mathbf{q}}^{\phi} + a E_{\mathbf{q}-\mathbf{k}}^{\chi}) t}{2}\right)}{(E_{\mathbf{k}}^{\phi} + E_{\mathbf{q}}^{\phi} + a E_{\mathbf{q}-\mathbf{k}}^{\chi})^2} f(E_{\mathbf{q}-\mathbf{k}}^{\chi}) \Biggr) 
\nonumber
\\
&&
+ \frac{\alpha_2^2}{4} \int_{\mathbf{k} \mathbf{q} \mathbf{l}} \frac{1}{2E_{\mathbf{k}}^{\phi} E_{\mathbf{k}+\mathbf{q}+\mathbf{l}}^{\phi} E_{\mathbf{q}}^{\chi}  E_{\mathbf{l}}^{\chi}} \Biggl( 
\sum_{a,b=\pm} 
\frac{\sin^2\left(\frac{(E_{\mathbf{k}}^{\phi} + E_{\mathbf{k}+\mathbf{q}+\mathbf{l}}^{\phi} + a E_{\mathbf{q}}^{\chi} + b E_{\mathbf{l}}^{\chi})t}{2} \right)}{2(E_{\mathbf{k}}^{\phi} + E_{\mathbf{k}+\mathbf{q}+\mathbf{l}}^{\phi} + a E_{\mathbf{q}}^{\chi} + b E_{\mathbf{l}}^{\chi})^2} f(E_{\mathbf{q}}^{\chi}) f(E_{\mathbf{l}}^{\chi}) 
\nonumber
\\
&&
\qquad + \frac{\sin^2\left(\frac{(E_{\mathbf{k}}^{\phi} + E_{\mathbf{k}+\mathbf{q}+\mathbf{l}}^{\phi} + E_{\mathbf{q}}^{\chi} + E_{\mathbf{l}}^{\chi})t}{2} \right)}{2(E_{\mathbf{k}}^{\phi} + E_{\mathbf{k}+\mathbf{q}+\mathbf{l}}^{\phi} + E_{\mathbf{q}}^{\chi} + E_{\mathbf{l}}^{\chi})^2}  + 
\sum_{a=\pm} 
\frac{\sin^2\left(\frac{(E_{\mathbf{k}}^{\phi} + E_{\mathbf{k}+\mathbf{q}+\mathbf{l}}^{\phi} + E_{\mathbf{q}}^{\chi} + a E_{\mathbf{l}}^{\chi})t}{2} \right)}{(E_{\mathbf{k}}^{\phi} + E_{\mathbf{k}+\mathbf{q}+\mathbf{l}}^{\phi} + E_{\mathbf{q}}^{\chi} + a E_{\mathbf{l}}^{\chi})^2} f(E_{\mathbf{l}}^{\chi}) \Biggr)\Biggr] 
\nonumber
\\ 
&&
- \frac{\mathrm{i} \alpha_2}{4} \left( \frac{1}{E_{\mathbf{p}}^{\phi}} - \frac{1}{E_{\mathbf{p'}}^{\phi}} \right) t \Delta^F_{zz} 
    - \frac{\alpha_2^2}{16} \Biggl[ \frac{1}{2} \left( \frac{1}{(E_{\mathbf{p}}^{\phi})^2} -\frac{2}{E_{\mathbf{p}}^{\phi}E_{\mathbf{p'}}^{\phi}} + \frac{1}{(E_{\mathbf{p'}}^{\phi})^2} \right)t^2 
\nonumber
\\ 
&& 
\qquad 
+ \left( \left( \frac{-\mathrm{i}}{2 (E_{\mathbf{p}}^{\phi})^3} t + \frac{1-e^{-2\mathrm{i} E_{\mathbf{p}}^{\phi}t}}{4(E_{\mathbf{p}}^{\phi})^4} \right) +(\mathbf{p} \longleftrightarrow \mathbf{p'}) ^* \right)  \Biggr] \Delta^\mathrm{F}_{zz} \Delta^\mathrm{F}_{z'z'} 
\nonumber
\\ 
&&
+ \mathrm{i} \alpha_1^2 M^2 \int_{\mathbf{q}} \Biggl[ \frac{1}{8 E_{\mathbf{p}}^{\phi} E_{\mathbf{q}}^{\phi} E_{\mathbf{p}- \mathbf{q}}^{\chi}} \Biggl(
\sum_{a=\pm} 
\frac{ \frac{\mathrm{i}}{a E_{\mathbf{p}}^{\phi} + E_{\mathbf{q}}^{\phi} + E_{\mathbf{p}-\mathbf{q}}^{\chi}} \left( 1 -e^{-\mathrm{i}(a E_{\mathbf{p}}^{\phi} + E_{\mathbf{q}}^{\phi} + E_{\mathbf{p}-\mathbf{q}}^{\chi})t} \right)+t}{a E_{\mathbf{p}}^{\phi} + E_{\mathbf{q}}^{\phi} + E_{\mathbf{p}-\mathbf{q}}^{\chi}}  
\nonumber
\\ 
&&
\qquad  
+ \sum_{a,b=\pm} \frac{ \frac{\mathrm{i}}{a E_{\mathbf{p}}^{\phi} + E_{\mathbf{q}}^{\phi} + b E_{\mathbf{p}-\mathbf{q}}^{\chi}} \left( 1 -e^{-\mathrm{i}(a E_{\mathbf{p}}^{\phi} + E_{\mathbf{q}}^{\phi} + b E_{\mathbf{p}-\mathbf{q}}^{\chi})t} \right) +t}{a E_{\mathbf{p}}^{\phi} + E_{\mathbf{q}}^{\phi} + b E_{\mathbf{p}-\mathbf{q}}^{\chi}} f(E_{\mathbf{p}-\mathbf{q}}^{\chi})\Biggr) + (\mathbf{p} \longleftrightarrow \mathbf{p'})^* \Biggr] 
\nonumber
\\
&&
+ \mathrm{i} \frac{\alpha_2^2}{2} \int_{\mathbf{k} \mathbf{q}} \Biggl[ \frac{1}{8 E_{\mathbf{p}}^{\phi} E_{\mathbf{p} - \mathbf{k} - \mathbf{q}}^{\phi} E_{\mathbf{k}}^{\chi} E_{\mathbf{q}}^{\chi}} \Biggl( 
\sum_{a=\pm}  \frac{\frac{\mathrm{i}}{a E_{\mathbf{p}}^{\phi} + E_{\mathbf{p} - \mathbf{k} - \mathbf{q}}^{\phi} + E_{\mathbf{k}}^{\chi} + E_{\mathbf{q}}^{\chi}} \left( 1 -e^{-\mathrm{i}(a E_{\mathbf{p}}^{\phi} + E_{\mathbf{p} - \mathbf{k} - \mathbf{q}}^{\phi} + E_{\mathbf{k}}^{\chi} + E_{\mathbf{q}}^{\chi})t} \right)  +t}{2(a E_{\mathbf{p}}^{\phi} + E_{\mathbf{p} - \mathbf{k} - \mathbf{q}}^{\phi} + E_{\mathbf{k}}^{\chi} + E_{\mathbf{q}}^{\chi})} 
\nonumber
\\ 
&&
\qquad+ 
\sum_{a,b,c=\pm}  \frac{\frac{\mathrm{i}}{a E_{\mathbf{p}}^{\phi} + E_{\mathbf{p} - \mathbf{k} - \mathbf{q}}^{\phi} + b E_{\mathbf{k}}^{\chi} + c E_{\mathbf{q}}^{\chi}} \left( 1 -e^{-\mathrm{i}(a E_{\mathbf{p}}^{\phi} + E_{\mathbf{p} - \mathbf{k} - \mathbf{q}}^{\phi} + b E_{\mathbf{k}}^{\chi} + c E_{\mathbf{q}}^{\chi})t} \right)+t}{2(a E_{\mathbf{p}}^{\phi} + E_{\mathbf{p} - \mathbf{k} - \mathbf{q}}^{\phi} + b E_{\mathbf{k}}^{\chi} + c E_{\mathbf{q}}^{\chi})} f(E_{\mathbf{k}}^{\chi})f(E_{\mathbf{q}}^{\chi}) 
\nonumber
\\ 
&&
\qquad  + 
\sum_{a,b=\pm}  \frac{\frac{\mathrm{i}}{a E_{\mathbf{p}}^{\phi} + E_{\mathbf{p} - \mathbf{k} - \mathbf{q}}^{\phi} + E_{\mathbf{k}}^{\chi} + b E_{\mathbf{q}}^{\chi}} \left( 1 -e^{-\mathrm{i}(a E_{\mathbf{p}}^{\phi} + E_{\mathbf{p} - \mathbf{k} - \mathbf{q}}^{\phi} + E_{\mathbf{k}}^{\chi} + b E_{\mathbf{q}}^{\chi})t} \right) +t}{a E_{\mathbf{p}}^{\phi} + E_{\mathbf{p} - \mathbf{k} - \mathbf{q}}^{\phi} + E_{\mathbf{k}}^{\chi} + b E_{\mathbf{q}}^{\chi}} f(E_{\mathbf{q}}^{\chi}) \Biggr) -(\mathbf{p} \longleftrightarrow \mathbf{p'})^* \Biggr]\Biggr\} 
\nonumber
\\
&&
+ \alpha_1^2 M^2 \int_{\mathbf{q}} \rho(\mathbf{p}-\mathbf{q}; \mathbf{p'} - \mathbf{q};0) e^{-\mathrm{i}(E_{\mathbf{p}}^{\phi} - E_{\mathbf{p'}}^{\phi}) t} \sum_{a=\pm} 
\left( e^{\mathrm{i}(E_{\mathbf{p}}^{\phi} - E_{\mathbf{p} -\mathbf{q}}^{\phi} - a E_{\mathbf{q}}^{\chi})t} -1 \right)
\nonumber
\\ 
&&
\qquad \times \left( e^{-\mathrm{i}(E_{\mathbf{p'}}^{\phi} - E_{\mathbf{p'} -\mathbf{q}}^{\phi} - a  E_{\mathbf{q}}^{\chi})t}-1 \right) \frac{a f(a E_{\mathbf{q}}^{\chi})}{8 E_{\mathbf{p} -\mathbf{q}}^{\phi} E_{\mathbf{p'}-\mathbf{q}}^{\phi} E_{\mathbf{q}}^{\chi} (E_{\mathbf{p}}^{\phi} - E_{\mathbf{p} -\mathbf{q}}^{\phi} - a E_{\mathbf{q}}^{\chi}) (E_{\mathbf{p'}}^{\phi} - E_{\mathbf{p'} -\mathbf{q}}^{\phi} - a E_{\mathbf{q}}^{\chi})}  
\nonumber
\\   
&&
+ \frac{\alpha_2^2}{2} \int_{\mathbf{k} \mathbf{q}} \rho(\mathbf{p}-\mathbf{k}-\mathbf{q}; \mathbf{p'} - \mathbf{k} - \mathbf{q};0) e^{-\mathrm{i}(E_{\mathbf{p}}^{\phi} - E_{\mathbf{p'}}^{\phi}) t} 
\nonumber
\\ 
&&
\qquad \times \sum_{a,b=\pm} 
\left( e^{\mathrm{i}(E_{\mathbf{p}}^{\phi} - E_{\mathbf{p} -\mathbf{k}-\mathbf{q}}^{\phi} - a E_{\mathbf{k}}^{\chi} - b E_{\mathbf{q}}^{\chi})t} -1 \right) \left( e^{-\mathrm{i}(E_{\mathbf{p'}}^{\phi} - E_{\mathbf{p'} -\mathbf{k} -\mathbf{q}}^{\phi} - a E_{\mathbf{k}}^{\chi} -b E_{\mathbf{q}}^{\chi})t}-1 \right)
\nonumber
\\ 
&&
\qquad \times \frac{ab f(a E_{\mathbf{k}}^{\chi})f(b E_{\mathbf{q}}^{\chi})}{16 E_{\mathbf{p} -\mathbf{k} -\mathbf{q}}^{\phi} E_{\mathbf{p'} -\mathbf{k} -\mathbf{q}}^{\phi} E_{\mathbf{k}}^{\chi} E_{\mathbf{q}}^{\chi} (E_{\mathbf{p}}^{\phi} - E_{\mathbf{p} -\mathbf{k}-\mathbf{q}}^{\phi} - a E_{\mathbf{k}}^{\chi} - b E_{\mathbf{q}}^{\chi}) (E_{\mathbf{p'}}^{\phi} - E_{\mathbf{p'} -\mathbf{k} -\mathbf{q}}^{\phi} - a E_{\mathbf{k}}^{\chi} -b E_{\mathbf{q}}^{\chi})}
\,\,\,.
\nonumber
\\
\end{eqnarray} 
 
\bibliography{DilatonOQD}

\providecommand{\href}[2]{#2}\begingroup\raggedright\begin{thebibliography}{100}

\bibitem{Fujii2003}
{Fujii, Yasunori and Maeda, Kei-ichi}, \emph{The Scalar-Tensor Theory of
  Gravitation}, Cambridge Monographs on Mathematical Physics. Cambridge
  University Press, 2003,
  \href{https://doi.org/10.1017/CBO9780511535093}{10.1017/CBO9780511535093}.

\bibitem{Clifton2011}
T.~Clifton, P.~G. Ferreira, A.~Padilla and C.~Skordis, \emph{Modified gravity
  and cosmology},
  \href{https://doi.org/https://doi.org/10.1016/j.physrep.2012.01.001}{\emph{Physics
  Reports} {\bfseries 513} (2012) 1}.

\bibitem{Joyce2014}
A.~Joyce, B.~Jain, J.~Khoury and M.~Trodden, \emph{{Beyond the Cosmological
  Standard Model}},
  \href{https://doi.org/10.1016/j.physrep.2014.12.002}{\emph{Phys. Rept.}
  {\bfseries 568} (2015) 1} [\href{https://arxiv.org/abs/1407.0059}{{\ttfamily
  1407.0059}}].

\bibitem{Dickey1994}
J.~O. Dickey, P.~L. Bender, J.~E. Faller, X.~X. Newhall, R.~L. Ricklefs, J.~G.
  Ries et~al., \emph{{Lunar Laser Ranging: A Continuing Legacy of the Apollo
  Program}}, \href{https://doi.org/10.1126/science.265.5171.482}{\emph{Science}
  {\bfseries 265} (1994) 482}.

\bibitem{Adelberger2003}
E.~Adelberger, B.~Heckel and A.~Nelson, \emph{{Tests of the Gravitational
  Inverse-Square Law}},
  \href{https://doi.org/10.1146/annurev.nucl.53.041002.110503}{\emph{Annual
  Review of Nuclear and Particle Science} {\bfseries 53} (2003) 77}.

\bibitem{Kapner2007}
D.~J. Kapner, T.~S. Cook, E.~G. Adelberger, J.~H. Gundlach, B.~R. Heckel, C.~D.
  Hoyle et~al., \emph{{Tests of the Gravitational Inverse-Square Law below the
  Dark-Energy Length Scale}},
  \href{https://doi.org/10.1103/PhysRevLett.98.021101}{\emph{Phys. Rev. Lett.}
  {\bfseries 98} (2007) 021101}.

\bibitem{BurrageSak}
C.~Burrage and J.~Sakstein, \emph{{Tests of Chameleon Gravity}},
  \href{https://doi.org/10.1007/s41114-018-0011-x}{\emph{Living Rev. Rel.}
  {\bfseries 21} (2018) 1} [\href{https://arxiv.org/abs/1709.09071}{{\ttfamily
  1709.09071}}].

\bibitem{Brax:2021wcv}
P.~Brax, S.~Casas, H.~Desmond and B.~Elder, \emph{{Testing Screened Modified
  Gravity}}, \href{https://doi.org/10.3390/universe8010011}{\emph{Universe}
  {\bfseries 8} (2021) 11} [\href{https://arxiv.org/abs/2201.10817}{{\ttfamily
  2201.10817}}].

\bibitem{Khoury2003}
J.~Khoury and A.~Weltman, \emph{{Chameleon cosmology}},
  \href{https://doi.org/10.1103/PhysRevD.69.044026}{\emph{Phys. Rev. D}
  {\bfseries 69} (2004) 044026}
  [\href{https://arxiv.org/abs/astro-ph/0309411}{{\ttfamily
  astro-ph/0309411}}].

\bibitem{Khoury20032}
J.~Khoury and A.~Weltman, \emph{{Chameleon fields: Awaiting surprises for tests
  of gravity in space}},
  \href{https://doi.org/10.1103/PhysRevLett.93.171104}{\emph{Phys. Rev. Lett.}
  {\bfseries 93} (2004) 171104}
  [\href{https://arxiv.org/abs/astro-ph/0309300}{{\ttfamily
  astro-ph/0309300}}].

\bibitem{Dehnen1992}
H.~Dehnen, H.~Frommert and F.~Ghaboussi, \emph{{Higgs field and a new scalar -
  tensor theory of gravity}},
  \href{https://doi.org/10.1007/BF00674344}{\emph{Int. J. Theor. Phys.}
  {\bfseries 31} (1992) 109}.

\bibitem{Gessner1992}
E.~Gessner, \emph{{A new scalar tensor theory for gravity and the flat rotation
  curves of spiral galaxies}},
  \href{https://doi.org/10.1007/BF00645239}{\emph{Astrophys. Space Sci.}
  {\bfseries 196} (1992) 29}.

\bibitem{Damour1994}
T.~Damour and A.~M. Polyakov, \emph{{The String dilaton and a least coupling
  principle}}, \href{https://doi.org/10.1016/0550-3213(94)90143-0}{\emph{Nucl.
  Phys. B} {\bfseries 423} (1994) 532}
  [\href{https://arxiv.org/abs/hep-th/9401069}{{\ttfamily hep-th/9401069}}].

\bibitem{Pietroni2005}
M.~Pietroni, \emph{Dark energy condensation},
  \href{https://doi.org/10.1103/PhysRevD.72.043535}{\emph{Phys. Rev. D}
  {\bfseries 72} (2005) 043535}.

\bibitem{Olive2008}
K.~A. Olive and M.~Pospelov, \emph{Environmental dependence of masses and
  coupling constants},
  \href{https://doi.org/10.1103/PhysRevD.77.043524}{\emph{Phys. Rev. D}
  {\bfseries 77} (2008) 043524}.

\bibitem{Brax2010}
P.~Brax, C.~van~de Bruck, A.-C. Davis and D.~Shaw, \emph{Dilaton and modified
  gravity}, \href{https://doi.org/10.1103/PhysRevD.82.063519}{\emph{Phys. Rev.
  D} {\bfseries 82} (2010) 063519}.

\bibitem{Hinterbichler2010}
K.~Hinterbichler and J.~Khoury, \emph{{Symmetron Fields: Screening Long-Range
  Forces Through Local Symmetry Restoration}},
  \href{https://doi.org/10.1103/PhysRevLett.104.231301}{\emph{Phys. Rev. Lett.}
  {\bfseries 104} (2010) 231301}
  [\href{https://arxiv.org/abs/1001.4525}{{\ttfamily 1001.4525}}].

\bibitem{Hinterbichler2011}
K.~Hinterbichler, J.~Khoury, A.~Levy and A.~Matas, \emph{{Symmetron
  Cosmology}}, \href{https://doi.org/10.1103/PhysRevD.84.103521}{\emph{Phys.
  Rev. D} {\bfseries 84} (2011) 103521}
  [\href{https://arxiv.org/abs/1107.2112}{{\ttfamily 1107.2112}}].

\bibitem{Burrage2016_2}
C.~Burrage, E.~J. Copeland and P.~Millington, \emph{{Radial acceleration
  relation from symmetron fifth forces}},
  \href{https://doi.org/10.1103/PhysRevD.95.064050}{\emph{Phys. Rev. D}
  {\bfseries 95} (2017) 064050}
  [\href{https://arxiv.org/abs/1610.07529}{{\ttfamily 1610.07529}}].

\bibitem{OHare:2018ayv}
C.~A.~J. O'Hare and C.~Burrage, \emph{{Stellar kinematics from the symmetron
  fifth force in the Milky Way disk}},
  \href{https://doi.org/10.1103/PhysRevD.98.064019}{\emph{Phys. Rev. D}
  {\bfseries 98} (2018) 064019}
  [\href{https://arxiv.org/abs/1805.05226}{{\ttfamily 1805.05226}}].

\bibitem{Burrage2018Sym}
C.~Burrage, E.~J. Copeland, C.~K\"ading and P.~Millington, \emph{{Symmetron
  scalar fields: Modified gravity, dark matter, or both?}},
  \href{https://doi.org/10.1103/PhysRevD.99.043539}{\emph{Phys. Rev. D}
  {\bfseries 99} (2019) 043539}
  [\href{https://arxiv.org/abs/1811.12301}{{\ttfamily 1811.12301}}].

\bibitem{Kading2023}
C.~K\"ading, \emph{{Lensing with generalized symmetrons}},
  \href{https://doi.org/10.3390/astronomy2020009}{\emph{Astronomy} {\bfseries
  2} (2023) 128} [\href{https://arxiv.org/abs/2304.05875}{{\ttfamily
  2304.05875}}].

\bibitem{Gasperini:2001pc}
M.~Gasperini, F.~Piazza and G.~Veneziano, \emph{{Quintessence as a runaway
  dilaton}}, \href{https://doi.org/10.1103/PhysRevD.65.023508}{\emph{Phys. Rev.
  D} {\bfseries 65} (2002) 023508}
  [\href{https://arxiv.org/abs/gr-qc/0108016}{{\ttfamily gr-qc/0108016}}].

\bibitem{Damour:2002nv}
T.~Damour, F.~Piazza and G.~Veneziano, \emph{{Violations of the equivalence
  principle in a dilaton runaway scenario}},
  \href{https://doi.org/10.1103/PhysRevD.66.046007}{\emph{Phys. Rev. D}
  {\bfseries 66} (2002) 046007}
  [\href{https://arxiv.org/abs/hep-th/0205111}{{\ttfamily hep-th/0205111}}].

\bibitem{Damour:2002mi}
T.~Damour, F.~Piazza and G.~Veneziano, \emph{{Runaway dilaton and equivalence
  principle violations}},
  \href{https://doi.org/10.1103/PhysRevLett.89.081601}{\emph{Phys. Rev. Lett.}
  {\bfseries 89} (2002) 081601}
  [\href{https://arxiv.org/abs/gr-qc/0204094}{{\ttfamily gr-qc/0204094}}].

\bibitem{Brax:2010gi}
P.~Brax, C.~van~de Bruck, A.-C. Davis and D.~Shaw, \emph{{The Dilaton and
  Modified Gravity}},
  \href{https://doi.org/10.1103/PhysRevD.82.063519}{\emph{Phys. Rev. D}
  {\bfseries 82} (2010) 063519}
  [\href{https://arxiv.org/abs/1005.3735}{{\ttfamily 1005.3735}}].

\bibitem{Brax:2011ja}
P.~Brax, C.~van~de Bruck, A.-C. Davis, B.~Li and D.~J. Shaw, \emph{{Nonlinear
  Structure Formation with the Environmentally Dependent Dilaton}},
  \href{https://doi.org/10.1103/PhysRevD.83.104026}{\emph{Phys. Rev. D}
  {\bfseries 83} (2011) 104026}
  [\href{https://arxiv.org/abs/1102.3692}{{\ttfamily 1102.3692}}].

\bibitem{Brax2022}
P.~Brax, H.~Fischer, C.~K\"ading and M.~Pitschmann, \emph{{The environment
  dependent dilaton in the laboratory and the solar system}},
  \href{https://doi.org/10.1140/epjc/s10052-022-10905-w}{\emph{Eur. Phys. J. C}
  {\bfseries 82} (2022) 934}
  [\href{https://arxiv.org/abs/2203.12512}{{\ttfamily 2203.12512}}].

\bibitem{Dvali2000}
G.~R. Dvali, G.~Gabadadze and M.~Porrati, \emph{{4-D gravity on a brane in 5-D
  Minkowski space}},
  \href{https://doi.org/10.1016/S0370-2693(00)00669-9}{\emph{Phys. Lett. B}
  {\bfseries 485} (2000) 208}
  [\href{https://arxiv.org/abs/hep-th/0005016}{{\ttfamily hep-th/0005016}}].

\bibitem{Nicolis2008}
A.~Nicolis, R.~Rattazzi and E.~Trincherini, \emph{{The Galileon as a local
  modification of gravity}},
  \href{https://doi.org/10.1103/PhysRevD.79.064036}{\emph{Phys. Rev. D}
  {\bfseries 79} (2009) 064036}
  [\href{https://arxiv.org/abs/0811.2197}{{\ttfamily 0811.2197}}].

\bibitem{Ali2012}
A.~Ali, R.~Gannouji, M.~W. Hossain and M.~Sami, \emph{{Light mass galileons:
  Cosmological dynamics, mass screening and observational constraints}},
  \href{https://doi.org/10.1016/j.physletb.2012.10.009}{\emph{Phys. Lett. B}
  {\bfseries 718} (2012) 5} [\href{https://arxiv.org/abs/1207.3959}{{\ttfamily
  1207.3959}}].

\bibitem{Burrage:2016bwy}
C.~Burrage and J.~Sakstein, \emph{{A Compendium of Chameleon Constraints}},
  \href{https://doi.org/10.1088/1475-7516/2016/11/045}{\emph{JCAP} {\bfseries
  11} (2016) 045} [\href{https://arxiv.org/abs/1609.01192}{{\ttfamily
  1609.01192}}].

\bibitem{Pokotilovski:2012xuk}
Y.~N. Pokotilovski, \emph{{Strongly coupled chameleon fields: Possible test
  with a neutron Lloyd's mirror interferometer}},
  \href{https://doi.org/10.1016/j.physletb.2013.01.022}{\emph{Phys. Lett. B}
  {\bfseries 719} (2013) 341}
  [\href{https://arxiv.org/abs/1203.5017}{{\ttfamily 1203.5017}}].

\bibitem{Pokotilovski:2013tma}
Y.~N. Pokotilovski, \emph{{Potential of the neutron Lloyd`s mirror
  interferometer for the search for new interactions}},
  \href{https://doi.org/10.1134/S106377611309001X}{\emph{J. Exp. Theor. Phys.}
  {\bfseries 116} (2013) 609}
  [\href{https://arxiv.org/abs/1311.4679}{{\ttfamily 1311.4679}}].

\bibitem{Burrage:2014oza}
C.~Burrage, E.~J. Copeland and E.~A. Hinds, \emph{{Probing Dark Energy with
  Atom Interferometry}},
  \href{https://doi.org/10.1088/1475-7516/2015/03/042}{\emph{JCAP} {\bfseries
  03} (2015) 042} [\href{https://arxiv.org/abs/1408.1409}{{\ttfamily
  1408.1409}}].

\bibitem{Hamilton:2015zga}
P.~Hamilton, M.~Jaffe, P.~Haslinger, Q.~Simmons, H.~M\"uller and J.~Khoury,
  \emph{{Atom-interferometry constraints on dark energy}},
  \href{https://doi.org/10.1126/science.aaa8883}{\emph{Science} {\bfseries 349}
  (2015) 849} [\href{https://arxiv.org/abs/1502.03888}{{\ttfamily
  1502.03888}}].

\bibitem{Lemmel:2015kwa}
H.~Lemmel, P.~Brax, A.~N. Ivanov, T.~Jenke, G.~Pignol, M.~Pitschmann et~al.,
  \emph{{Neutron Interferometry constrains dark energy chameleon fields}},
  \href{https://doi.org/10.1016/j.physletb.2015.02.063}{\emph{Phys. Lett. B}
  {\bfseries 743} (2015) 310}
  [\href{https://arxiv.org/abs/1502.06023}{{\ttfamily 1502.06023}}].

\bibitem{Burrage:2015lya}
C.~Burrage and E.~J. Copeland, \emph{{Using Atom Interferometry to Detect Dark
  Energy}}, \href{https://doi.org/10.1080/00107514.2015.1060058}{\emph{Contemp.
  Phys.} {\bfseries 57} (2016) 164}
  [\href{https://arxiv.org/abs/1507.07493}{{\ttfamily 1507.07493}}].

\bibitem{Elder:2016yxm}
B.~Elder, J.~Khoury, P.~Haslinger, M.~Jaffe, H.~M\"uller and P.~Hamilton,
  \emph{{Chameleon Dark Energy and Atom Interferometry}},
  \href{https://doi.org/10.1103/PhysRevD.94.044051}{\emph{Phys. Rev. D}
  {\bfseries 94} (2016) 044051}
  [\href{https://arxiv.org/abs/1603.06587}{{\ttfamily 1603.06587}}].

\bibitem{Ivanov:2016rfs}
A.~N. Ivanov, G.~Cronenberg, R.~H\"ollwieser, M.~Pitschmann, T.~Jenke,
  M.~Wellenzohn et~al., \emph{{Exact solution for chameleon field, self-coupled
  through the Ratra-Peebles potential with $n=1$ and confined between two
  parallel plates}},
  \href{https://doi.org/10.1103/PhysRevD.94.085005}{\emph{Phys. Rev. D}
  {\bfseries 94} (2016) 085005}
  [\href{https://arxiv.org/abs/1606.06867}{{\ttfamily 1606.06867}}].

\bibitem{Burrage:2016rkv}
C.~Burrage, A.~Kuribayashi-Coleman, J.~Stevenson and B.~Thrussell,
  \emph{{Constraining symmetron fields with atom interferometry}},
  \href{https://doi.org/10.1088/1475-7516/2016/12/041}{\emph{JCAP} {\bfseries
  12} (2016) 041} [\href{https://arxiv.org/abs/1609.09275}{{\ttfamily
  1609.09275}}].

\bibitem{Jaffe:2016fsh}
M.~Jaffe, P.~Haslinger, V.~Xu, P.~Hamilton, A.~Upadhye, B.~Elder et~al.,
  \emph{{Testing sub-gravitational forces on atoms from a miniature, in-vacuum
  source mass}}, \href{https://doi.org/10.1038/nphys4189}{\emph{Nature Phys.}
  {\bfseries 13} (2017) 938}
  [\href{https://arxiv.org/abs/1612.05171}{{\ttfamily 1612.05171}}].

\bibitem{Brax:2017hna}
P.~Brax and M.~Pitschmann, \emph{{Exact solutions to nonlinear symmetron
  theory: One- and two-mirror systems}},
  \href{https://doi.org/10.1103/PhysRevD.97.064015}{\emph{Phys. Rev. D}
  {\bfseries 97} (2018) 064015}
  [\href{https://arxiv.org/abs/1712.09852}{{\ttfamily 1712.09852}}].

\bibitem{Sabulsky:2018jma}
D.~O. Sabulsky, I.~Dutta, E.~A. Hinds, B.~Elder, C.~Burrage and E.~J. Copeland,
  \emph{{Experiment to detect dark energy forces using atom interferometry}},
  \href{https://doi.org/10.1103/PhysRevLett.123.061102}{\emph{Phys. Rev. Lett.}
  {\bfseries 123} (2019) 061102}
  [\href{https://arxiv.org/abs/1812.08244}{{\ttfamily 1812.08244}}].

\bibitem{Brax:2018iyo}
P.~Brax, C.~Burrage and A.-C. Davis, \emph{{Laboratory constraints}},
  \href{https://doi.org/10.1142/S0218271818480097}{\emph{Int. J. Mod. Phys. D}
  {\bfseries 27} (2018) 1848009}.

\bibitem{Cronenberg:2018qxf}
G.~Cronenberg, P.~Brax, H.~Filter, P.~Geltenbort, T.~Jenke, G.~Pignol et~al.,
  \emph{{Acoustic Rabi oscillations between gravitational quantum states and
  impact on symmetron dark energy}},
  \href{https://doi.org/10.1038/s41567-018-0205-x}{\emph{Nature Phys.}
  {\bfseries 14} (2018) 1022}
  [\href{https://arxiv.org/abs/1902.08775}{{\ttfamily 1902.08775}}].

\bibitem{Hartley2019}
D.~Hartley, C.~K\"ading, R.~Howl and I.~Fuentes, \emph{{Quantum-enhanced
  screened dark energy detection}},
  \href{https://arxiv.org/abs/1909.02272}{{\ttfamily 1909.02272}}.

\bibitem{Pitschmann:2020ejb}
M.~Pitschmann, \emph{{Exact solutions to nonlinear symmetron theory: One- and
  two-mirror systems. II.}},
  \href{https://doi.org/10.1103/PhysRevD.103.084013}{\emph{Phys. Rev. D}
  {\bfseries 103} (2021) 084013}
  [\href{https://arxiv.org/abs/2012.12752}{{\ttfamily 2012.12752}}].

\bibitem{Brax2018quantch}
P.~Brax and S.~Fichet, \emph{{Quantum Chameleons}},
  \href{https://doi.org/10.1103/PhysRevD.99.104049}{\emph{Phys. Rev. D}
  {\bfseries 99} (2019) 104049}
  [\href{https://arxiv.org/abs/1809.10166}{{\ttfamily 1809.10166}}].

\bibitem{Burrage2018}
C.~Burrage, C.~K\"ading, P.~Millington and J.~Min\'a\v{r}, \emph{{Open quantum
  dynamics induced by light scalar fields}},
  \href{https://doi.org/10.1103/PhysRevD.100.076003}{\emph{Phys. Rev. D}
  {\bfseries 100} (2019) 076003}
  [\href{https://arxiv.org/abs/1812.08760}{{\ttfamily 1812.08760}}].

\bibitem{Burrage2019}
C.~Burrage, C.~K\"ading, P.~Millington and J.~Min\'a\v{r}, \emph{{Influence
  functionals, decoherence and conformally coupled scalars}},
  \href{https://doi.org/10.1088/1742-6596/1275/1/012041}{\emph{J. Phys. Conf.
  Ser.} {\bfseries 1275} (2019) 012041}
  [\href{https://arxiv.org/abs/1902.09607}{{\ttfamily 1902.09607}}].

\bibitem{Kading2019}
C.~K\"ading, \emph{{Astro- and Quantum Physical Tests of Screened Scalar
  Fields}}, Ph.D. thesis, University of Nottingham, Nottingham NG7 2RD, UK, 10,
  2019.
\newblock \href{https://arxiv.org/abs/1910.05738}{{\ttfamily 1910.05738}}.

\bibitem{Hartley2018}
D.~Hartley, C.~K\"ading, R.~Howl and I.~Fuentes, \emph{{Quantum simulation of
  dark energy candidates}},
  \href{https://doi.org/10.1103/PhysRevD.99.105002}{\emph{Phys. Rev. D}
  {\bfseries 99} (2019) 105002}
  [\href{https://arxiv.org/abs/1811.06927}{{\ttfamily 1811.06927}}].

\bibitem{Breuer2002}
H.-P. Breuer and F.~Petruccione, \emph{The Theory of Open Quantum Systems}.
  Oxford University Press, Oxford, 2002.

\bibitem{Schlosshauer}
M.~Schlosshauer, \emph{Decoherence and the Quantum-To-Classical Transition}.
  Springer-Verlag Berlin Heidelberg, 2007.

\bibitem{Schwinger}
J.~S. Schwinger, \emph{{Brownian Motion of a Quantum Oscillator}},
  \href{https://doi.org/10.1063/1.1703727}{\emph{J. Math. Phys.} {\bfseries 2}
  (1961) 407}.

\bibitem{Keldysh}
L.~V. Keldysh, \emph{{Diagram technique for nonequilibrium processes}},
  {\emph{Zh. Eksp. Teor. Fiz.} {\bfseries 47} (1964) 1515}.

\bibitem{Feynman}
R.~P. Feynman and F.~L. Vernon, \emph{The theory of a general quantum system
  interacting with a linear dissipative system}, {\emph{Annals of physics}
  {\bfseries 24} (1963) 118}.

\bibitem{Kading2022x}
C.~K\"ading and M.~Pitschmann, \emph{{New method for directly computing reduced
  density matrices}},
  \href{https://doi.org/10.1103/PhysRevD.107.016005}{\emph{Phys. Rev. D}
  {\bfseries 107} (2023) 016005}
  [\href{https://arxiv.org/abs/2204.08829}{{\ttfamily 2204.08829}}].

\bibitem{Kading2022_2}
C.~K\"ading and M.~Pitschmann, \emph{{Density Matrix Formalism for Interacting
  Quantum Fields}},
  \href{https://doi.org/10.3390/universe8110601}{\emph{Universe} {\bfseries 8}
  (2022) 601} [\href{https://arxiv.org/abs/2210.06991}{{\ttfamily
  2210.06991}}].

\bibitem{Carmichael}
H.~Carmichael, \emph{An Open Systems Approach to Quantum Optics: Lectures
  Presented at the Universit{\'e} Libre de Bruxelles, October 28 to November 4,
  1991}. Springer Berlin Heidelberg, 1993.

\bibitem{Gardiner2004}
C.~Gardiner and P.~Zoller, \emph{Quantum Noise: A Handbook of Markovian and
  Non-Markovian Quantum Stochastic Methods with Applications to Quantum
  Optics}, Springer Series in Synergetics. Springer, 2004.

\bibitem{Walls2008}
D.~Walls and G.~Milburn, \emph{Quantum Optics}. Springer Berlin Heidelberg,
  2008.

\bibitem{Aolita2015}
L.~Aolita, F.~de~Melo and L.~Davidovich, \emph{Open-system dynamics of
  entanglement:a key issues review},
  \href{https://doi.org/10.1088/0034-4885/78/4/042001}{\emph{Reports on
  Progress in Physics} {\bfseries 78} (2015) 042001}.

\bibitem{Goold2016}
J.~Goold, M.~Huber, A.~Riera, L.~d. Rio and P.~Skrzypczyk, \emph{The role of
  quantum information in thermodynamics—a topical review},
  \href{https://doi.org/10.1088/1751-8113/49/14/143001}{\emph{Journal of
  Physics A: Mathematical and Theoretical} {\bfseries 49} (2016) 143001}.

\bibitem{Werner2016}
A.~Werner, D.~Jaschke, P.~Silvi, M.~Kliesch, T.~Calarco, J.~Eisert et~al.,
  \emph{{Positive Tensor Network Approach for Simulating Open Quantum Many-Body
  Systems}},
  \href{https://doi.org/10.1103/physrevlett.116.237201}{\emph{Physical Review
  Letters} {\bfseries 116} (2016) }.

\bibitem{Huber2020}
J.~Huber, P.~Kirton, S.~Rotter and P.~Rabl, \emph{{Emergence of PT-symmetry
  breaking in open quantum systems}},
  \href{https://doi.org/10.21468/scipostphys.9.4.052}{\emph{SciPost Physics}
  {\bfseries 9} (2020) }.

\bibitem{Calzetta2008}
E.~A. Calzetta and B.-L. Hu, \emph{Nonequilibrium Quantum Field Theory}.
  Cambridge University Press, Cambridge UK, 2008.

\bibitem{Koksma2010}
J.~F. Koksma, T.~Prokopec and M.~G. Schmidt, \emph{Decoherence in an
  interacting quantum field theory: The vacuum case},
  \href{https://doi.org/10.1103/PhysRevD.81.065030}{\emph{Phys. Rev. D}
  {\bfseries 81} (2010) 065030}.

\bibitem{Koksma2011}
J.~F. Koksma, T.~Prokopec and M.~G. Schmidt, \emph{Decoherence in an
  interacting quantum field theory: Thermal case},
  \href{https://doi.org/10.1103/PhysRevD.83.085011}{\emph{Phys. Rev. D}
  {\bfseries 83} (2011) 085011}.

\bibitem{Sieberer2016}
L.~M. Sieberer, M.~Buchhold and S.~Diehl, \emph{Keldysh field theory for driven
  open quantum systems}, {\emph{Reports on Progress in Physics} {\bfseries 79}
  (2016) 096001}.

\bibitem{Marino2016}
J.~Marino and S.~Diehl, \emph{Quantum dynamical field theory for nonequilibrium
  phase transitions in driven open systems},
  \href{https://doi.org/10.1103/PhysRevB.94.085150}{\emph{Phys. Rev. B}
  {\bfseries 94} (2016) 085150}.

\bibitem{Baidya2017}
A.~Baidya, C.~Jana, R.~Loganayagam and A.~Rudra, \emph{{Renormalization in open
  quantum field theory. Part I. Scalar field theory}},
  \href{https://doi.org/10.1007/JHEP11(2017)204}{\emph{JHEP} {\bfseries 11}
  (2017) 204} [\href{https://arxiv.org/abs/1704.08335}{{\ttfamily
  1704.08335}}].

\bibitem{Nagy2020}
S.~Nagy and J.~Polonyi, \emph{{Renormalizing Open Quantum Field Theories}},
  \href{https://doi.org/10.3390/universe8020127}{\emph{Universe} {\bfseries 8}
  (2022) 127} [\href{https://arxiv.org/abs/2012.13811}{{\ttfamily
  2012.13811}}].

\bibitem{Jana2021}
C.~Jana, \emph{{Aspects of open quantum field theory}}, Ph.D. thesis, Tata
  Inst., 2021.

\bibitem{Fogedby2022}
H.~C. Fogedby, \emph{{Field-theoretical approach to open quantum systems and
  the Lindblad equation}},
  \href{https://doi.org/10.1103/PhysRevA.106.022205}{\emph{Phys. Rev. A}
  {\bfseries 106} (2022) 022205}
  [\href{https://arxiv.org/abs/2202.05203}{{\ttfamily 2202.05203}}].

\bibitem{Lombardo1}
F.~Lombardo and F.~D. Mazzitelli, \emph{Coarse graining and decoherence in
  quantum field theory},
  \href{https://doi.org/10.1103/PhysRevD.53.2001}{\emph{Phys. Rev. D}
  {\bfseries 53} (1996) 2001}.

\bibitem{Lombardo2}
F.~C. Lombardo and D.~L. Nacir, \emph{Decoherence during inflation: The
  generation of classical inhomogeneities},
  \href{https://doi.org/10.1103/PhysRevD.72.063506}{\emph{Phys. Rev. D}
  {\bfseries 72} (2005) 063506}.

\bibitem{Lombardo3}
F.~C. Lombardo, \emph{{Influence functional approach to decoherence during
  inflation}},
  \href{https://doi.org/10.1590/S0103-97332005000300005}{\emph{Braz. J. Phys.}
  {\bfseries 35} (2005) 391}
  [\href{https://arxiv.org/abs/gr-qc/0412069}{{\ttfamily gr-qc/0412069}}].

\bibitem{Boyanovsky1}
D.~Boyanovsky, \emph{{Effective field theory during inflation: Reduced density
  matrix and its quantum master equation}},
  \href{https://doi.org/10.1103/PhysRevD.92.023527}{\emph{Phys. Rev.}
  {\bfseries D92} (2015) 023527}
  [\href{https://arxiv.org/abs/1506.07395}{{\ttfamily 1506.07395}}].

\bibitem{Boyanovsky2}
D.~Boyanovsky, \emph{{Effective field theory during inflation. II. Stochastic
  dynamics and power spectrum suppression}},
  \href{https://doi.org/10.1103/PhysRevD.93.043501}{\emph{Phys. Rev.}
  {\bfseries D93} (2016) 043501}
  [\href{https://arxiv.org/abs/1511.06649}{{\ttfamily 1511.06649}}].

\bibitem{Boyanovsky3}
D.~Boyanovsky, \emph{{Fermionic influence on inflationary fluctuations}},
  \href{https://doi.org/10.1103/PhysRevD.93.083507}{\emph{Phys. Rev.}
  {\bfseries D93} (2016) 083507}
  [\href{https://arxiv.org/abs/1602.05609}{{\ttfamily 1602.05609}}].

\bibitem{Boyanovsky4}
D.~Boyanovsky, \emph{{Imprint of entanglement entropy in the power spectrum of
  inflationary fluctuations}},
  \href{https://doi.org/10.1103/PhysRevD.98.023515}{\emph{Phys. Rev.}
  {\bfseries D98} (2018) 023515}
  [\href{https://arxiv.org/abs/1804.07967}{{\ttfamily 1804.07967}}].

\bibitem{Burgess2015}
C.~P. Burgess, R.~Holman, G.~Tasinato and M.~Williams, \emph{{EFT beyond the
  horizon: stochastic inflation and how primordial quantum fluctuations go
  classical}}, \href{https://doi.org/10.1007/JHEP03(2015)090}{\emph{Journal of
  High Energy Physics} {\bfseries 2015} (2015) 90}.

\bibitem{Hollowood}
T.~J. Hollowood and J.~I. McDonald, \emph{Decoherence, discord, and the quantum
  master equation for cosmological perturbations},
  \href{https://doi.org/10.1103/PhysRevD.95.103521}{\emph{Phys. Rev. D}
  {\bfseries 95} (2017) 103521}.

\bibitem{Binder2021}
T.~Binder, K.~Mukaida, B.~Scheihing-Hitschfeld and X.~Yao, \emph{{Non-Abelian
  electric field correlator at NLO for dark matter relic abundance and
  quarkonium transport}},
  \href{https://doi.org/10.1007/JHEP01(2022)137}{\emph{JHEP} {\bfseries 01}
  (2022) 137} [\href{https://arxiv.org/abs/2107.03945}{{\ttfamily
  2107.03945}}].

\bibitem{Brahma:2022yxu}
S.~Brahma, A.~Berera and J.~Calder\'on-Figueroa, \emph{{Quantum corrections to
  the primordial tensor spectrum: open EFTs \& Markovian decoupling of UV
  modes}}, \href{https://doi.org/10.1007/JHEP08(2022)225}{\emph{JHEP}
  {\bfseries 08} (2022) 225}
  [\href{https://arxiv.org/abs/2206.05797}{{\ttfamily 2206.05797}}].

\bibitem{Brahma:2021mng}
S.~Brahma, A.~Berera and J.~Calder\'on-Figueroa, \emph{{Universal signature of
  quantum entanglement across cosmological distances}},
  \href{https://doi.org/10.1088/1361-6382/aca066}{\emph{Class. Quant. Grav.}
  {\bfseries 39} (2022) 245002}
  [\href{https://arxiv.org/abs/2107.06910}{{\ttfamily 2107.06910}}].

\bibitem{Colas:2022hlq}
T.~Colas, J.~Grain and V.~Vennin, \emph{{Benchmarking the cosmological master
  equations}},
  \href{https://doi.org/10.1140/epjc/s10052-022-11047-9}{\emph{Eur. Phys. J. C}
  {\bfseries 82} (2022) 1085}
  [\href{https://arxiv.org/abs/2209.01929}{{\ttfamily 2209.01929}}].

\bibitem{Yu2008}
H.~W. Yu, J.~Zhang, H.-w. Yu and J.-l. Zhang, \emph{{Understanding Hawking
  radiation in the framework of open quantum systems}},
  \href{https://doi.org/10.1103/PhysRevD.77.029904}{\emph{Phys. Rev. D}
  {\bfseries 77} (2008) 024031}
  [\href{https://arxiv.org/abs/0806.3602}{{\ttfamily 0806.3602}}].

\bibitem{Lombardo2012}
F.~C. Lombardo and G.~J. Turiaci, \emph{{Dynamics of an Acoustic Black Hole as
  an Open Quantum System}},
  \href{https://doi.org/10.1103/PhysRevD.87.084028}{\emph{Phys. Rev. D}
  {\bfseries 87} (2013) 084028}
  [\href{https://arxiv.org/abs/1208.0198}{{\ttfamily 1208.0198}}].

\bibitem{Jana2020}
C.~Jana, R.~Loganayagam and M.~Rangamani, \emph{{Open quantum systems and
  Schwinger-Keldysh holograms}},
  \href{https://doi.org/10.1007/JHEP07(2020)242}{\emph{JHEP} {\bfseries 07}
  (2020) 242} [\href{https://arxiv.org/abs/2004.02888}{{\ttfamily
  2004.02888}}].

\bibitem{Agarwal2020}
K.~Agarwal and N.~Bao, \emph{Toy model for decoherence in the black hole
  information problem},
  \href{https://doi.org/10.1103/PhysRevD.102.086017}{\emph{Phys. Rev. D}
  {\bfseries 102} (2020) 086017}.

\bibitem{Kaplanek2020}
G.~Kaplanek and C.~P. Burgess, \emph{{Qubits on the Horizon: Decoherence and
  Thermalization near Black Holes}},
  \href{https://doi.org/10.1007/JHEP01(2021)098}{\emph{JHEP} {\bfseries 01}
  (2021) 098} [\href{https://arxiv.org/abs/2007.05984}{{\ttfamily
  2007.05984}}].

\bibitem{Burgess2021}
C.~P. Burgess, R.~Holman and G.~Kaplanek, \emph{{Quantum Hotspots: Mean Fields,
  Open EFTs, Nonlocality and Decoherence Near Black Holes}},
  \href{https://arxiv.org/abs/2106.10804}{{\ttfamily 2106.10804}}.

\bibitem{Kaplanek2021}
G.~Kaplanek, C.~P. Burgess and R.~Holman, \emph{{Qubit heating near a
  hotspot}}, \href{https://doi.org/10.1007/JHEP08(2021)132}{\emph{JHEP}
  {\bfseries 08} (2021) 132}
  [\href{https://arxiv.org/abs/2106.10803}{{\ttfamily 2106.10803}}].

\bibitem{Brambilla1}
N.~Brambilla, M.~A. Escobedo, J.~Soto and A.~Vairo, \emph{{Quarkonium
  suppression in heavy-ion collisions: an open quantum system approach}},
  \href{https://doi.org/10.1103/PhysRevD.96.034021}{\emph{Phys. Rev.}
  {\bfseries D96} (2017) 034021}
  [\href{https://arxiv.org/abs/1612.07248}{{\ttfamily 1612.07248}}].

\bibitem{Brambilla2}
N.~Brambilla, M.~A. Escobedo, J.~Soto and A.~Vairo, \emph{{Heavy quarkonium
  suppression in a fireball}},
  \href{https://doi.org/10.1103/PhysRevD.97.074009}{\emph{Phys. Rev.}
  {\bfseries D97} (2018) 074009}
  [\href{https://arxiv.org/abs/1711.04515}{{\ttfamily 1711.04515}}].

\bibitem{Yao2018}
X.~Yao and T.~Mehen, \emph{{Quarkonium in-medium transport equation derived
  from first principles}},
  \href{https://doi.org/10.1103/PhysRevD.99.096028}{\emph{Phys. Rev. D}
  {\bfseries 99} (2019) 096028}
  [\href{https://arxiv.org/abs/1811.07027}{{\ttfamily 1811.07027}}].

\bibitem{Yao2020}
X.~Yao and T.~Mehen, \emph{{Quarkonium Semiclassical Transport in Quark-Gluon
  Plasma: Factorization and Quantum Correction}},
  \href{https://doi.org/10.1007/JHEP02(2021)062}{\emph{JHEP} {\bfseries 02}
  (2021) 062} [\href{https://arxiv.org/abs/2009.02408}{{\ttfamily
  2009.02408}}].

\bibitem{Akamatsu2020}
Y.~Akamatsu, \emph{{Quarkonium in quark\textendash{}gluon plasma: Open quantum
  system approaches re-examined}},
  \href{https://doi.org/10.1016/j.ppnp.2021.103932}{\emph{Prog. Part. Nucl.
  Phys.} {\bfseries 123} (2022) 103932}
  [\href{https://arxiv.org/abs/2009.10559}{{\ttfamily 2009.10559}}].

\bibitem{DeJong2020}
W.~A. De~Jong, M.~Metcalf, J.~Mulligan, M.~P\l{}osko\'n, F.~Ringer and X.~Yao,
  \emph{{Quantum simulation of open quantum systems in heavy-ion collisions}},
  \href{https://doi.org/10.1103/PhysRevD.104.L051501}{\emph{Phys. Rev. D}
  {\bfseries 104} (2021) 051501}
  [\href{https://arxiv.org/abs/2010.03571}{{\ttfamily 2010.03571}}].

\bibitem{Yao2021}
X.~Yao, \emph{{Open quantum systems for quarkonia}},
  \href{https://doi.org/10.1142/S0217751X21300106}{\emph{Int. J. Mod. Phys. A}
  {\bfseries 36} (2021) 2130010}
  [\href{https://arxiv.org/abs/2102.01736}{{\ttfamily 2102.01736}}].

\bibitem{Brambilla2021}
N.~Brambilla, M.~A. Escobedo, M.~Strickland, A.~Vairo, P.~Vander~Griend and
  J.~H. Weber, \emph{{Bottomonium production in heavy-ion collisions using
  quantum trajectories: Differential observables and momentum anisotropy}},
  \href{https://doi.org/10.1103/PhysRevD.104.094049}{\emph{Phys. Rev. D}
  {\bfseries 104} (2021) 094049}
  [\href{https://arxiv.org/abs/2107.06222}{{\ttfamily 2107.06222}}].

\bibitem{Griend2021}
P.~V. Griend, \emph{{Bottomonium observables in an open quantum system using
  the quantum trajectories method}},
  \href{https://doi.org/10.1051/epjconf/202225805005}{\emph{EPJ Web Conf.}
  {\bfseries 258} (2022) 05005}
  [\href{https://arxiv.org/abs/2111.13520}{{\ttfamily 2111.13520}}].

\bibitem{Yao2022}
X.~Yao, \emph{{Quarkonium Suppression in the Open Quantum System Approach}},
  in \emph{{19th International Conference on Hadron Spectroscopy and
  Structure}}, 1, 2022, \href{https://arxiv.org/abs/2201.07702}{{\ttfamily
  2201.07702}}.

\bibitem{Wick}
G.~C. Wick, \emph{The evaluation of the collision matrix},
  \href{https://doi.org/10.1103/PhysRev.80.268}{\emph{Phys. Rev.} {\bfseries
  80} (1950) 268}.

\bibitem{bcca-m22}
B.~Barrett, G.~Condon, L.~Chichet, L.~Antoni-Micollier, R.~Arguel, M.~Rabault
  et~al., \emph{Testing the universality of free fall using correlated
  39k–87rb atom interferometers}, {\emph{AVS Quantum Science} {\bfseries 4}
  (2022) 014401}.

\bibitem{eymkl15}
B.~Estey, C.~Yu, H.~Müller, P.-C. Kuan and S.-Y. Lan, \emph{{High-Resolution
  Atom Interferometers with Suppressed Diffraction Phases}}, {\emph{Phys. Rev.
  Lett.} {\bfseries 115} (2015) 083002}.

\bibitem{hydrogen}
{\relax National Center for Biotechnology Information}, ``{PubChem Compound
  Summary for CID 783, Hydrogen}.''
  \url{https://pubchem.ncbi.nlm.nih.gov/compound/Hydrogen}, 2023.

\bibitem{lowdens}
G.~Gabrielse, X.~Fei, L.~A. Orozco, R.~L. Tjoelker, J.~Haas, H.~Kalinowsky
  et~al., \emph{Thousandfold improvement in the measured antiproton mass},
  \href{https://doi.org/10.1103/PhysRevLett.65.1317}{\emph{Phys. Rev. Lett.}
  {\bfseries 65} (1990) 1317}.

\bibitem{rubidium}
{\relax National Center for Biotechnology Information}, ``{PubChem Compound
  Summary for CID 5357696, Rubidium}.''
  \url{https://pubchem.ncbi.nlm.nih.gov/compound/Rubidium}, 2023.

\bibitem{warmvap}
G.~W. Biedermann, H.~J. McGuinness, A.~V. Rakholia, Y.-Y. Jau, D.~R. Wheeler,
  J.~D. Sterk et~al., \emph{{Atom Interferometry in a Warm Vapor}},
  \href{https://doi.org/10.1103/PhysRevLett.118.163601}{\emph{Phys. Rev. Lett.}
  {\bfseries 118} (2017) 163601}.

\end{thebibliography}\endgroup
\bibliographystyle{JHEP}

\end{document}